\documentclass[pdflatex,sn-mathphys-num]{sn-jnl}
\usepackage{hyperref}

\usepackage{graphicx}%
\usepackage{multirow}%
\usepackage{amsmath,amssymb,amsfonts}%
\usepackage{amsthm}%
\usepackage{mathrsfs}%
\usepackage[title]{appendix}%
\usepackage{xcolor}%
\usepackage{textcomp}%
\usepackage{manyfoot}%
\usepackage{enumitem}
\usepackage{booktabs}%
\usepackage{algorithm}%
\usepackage{algorithmicx}%
\usepackage{algpseudocode}%
\usepackage{listings}%
\usepackage{comment}
\usepackage{marginnote}
\usepackage{tabularx}
\usepackage{graphicx}
\usepackage{datatool}
\usepackage{pdflscape}
\usepackage{makecell}

\usepackage[most]{tcolorbox}

\DTLloaddb[noheader, keys={variable,value}]{mydata}{Data/Data.dat}

\newcommand{\GetValue}[1]{\DTLfetch{mydata}{variable}{#1}{value}}

\usepackage{bbm}
\newboolean{showcomments}
\setboolean{showcomments}{true}
\ifthenelse{\boolean{showcomments}}
{\newcommand{\nb}[2]{
		\fcolorbox{black}{yellow}{\bfseries\sffamily\scriptsize#1}
		{\sf\small\textit{#2}}
	}
	
}
{\newcommand{\nb}[2]{}
	
}

\theoremstyle{thmstyleone}%
\theoremstyle{thmstyletwo}%

\theoremstyle{thmstylethree}%

\begin{document}
\begin{filecontents*}{Data/Data.dat}
FinalPPAITeration2, 90\%
Total_Sources, 8482
\end{filecontents*}

\title{Accelerating Discovery: Rapid Literature Screening with LLMs}

%
%
%


\author*[1]{\fnm{Santiago} \sur{Matalonga}}\email{santiago.matalonga@uws.ac.uk}

\author[2]{\fnm{Domenico} \sur{Amalfitano}}\email{domenico.amalfitano@unina.it}
\equalcont{These authors contributed equally to this work.}

\author[3]{\fnm{Jean Carlo} \sur{Rossa Hauck}}\email{jean.hauck@ufsc.br}
\equalcont{These authors contributed equally to this work.}

\author[4]{\fnm{Martín} \sur{Solari}}\email{martin.solari@ort.edu.uy}
\equalcont{These authors contributed equally to this work.}

\author[5]{\fnm{Guilherme} \sur{H. Travassos}}\email{ght@cos.ufrj.br}
\equalcont{These authors contributed equally to this work.}

\affil*[1]{\orgdiv{Computing, Engineering and Physical Science}, \orgname{University of the West of Scotland}, \orgaddress{\street{High Street}, \city{Paisley}, \postcode{PA1 2BE}, \state{Renfrewshire}, \country{United Kingdom}}}

\affil[2]{\orgdiv{Department of Electrical Engineering and Information Technology (DIETI)}, \orgname{University of Naples Federico II}, \orgaddress{\street{Via Claudio, 21}, \city{Napoli}, \postcode{80125}, \state{NA}, \country{Italy}}}

\affil[3]{\orgdiv{Departamento de Informática e Estatística}, \orgname{Universidade Federal de Santa Catarina}, \orgaddress{\street{R. Rua Delfino Conti, s/n}, \city{Florianopolis}, \postcode{88040-370}, \state{Santa Catharina}, \country{Brasil}}}

\affil[4]{\orgdiv{Facultad de Ingeniería}, \orgname{Universidad ORT Uruguay}, \orgaddress{\street{Cuareim 1451}, \city{Montevideo}, \postcode{11100}, \country{Uruguay}}}

\affil[5]{\orgdiv{Programa de Engenharia de Sistemas e Computação}, \orgname{Univesidade Federal do Rio de Janheiro}, \orgaddress{\street{Avenida Horácio Macedo 2030, Bloco H, Centro de Tecnologia}, \city{Rio de Janheiro}, \postcode{21941-914}, \state{RJ}, \country{Brasil}}}


\abstract{
\textbf{Background:} Conducting Multi-Vocal Literature Reviews (MVLRs) is often time and effort-intensive. Researchers must review and filter a large number of unstructured sources, which frequently contain sparse information and are unlikely to be included in the final study. Our experience conducting an MVLR on Context-Aware Software Systems (CASS) Testing in the avionics domain exemplified this challenge, with over 8,000 highly heterogeneous documents requiring review. Therefore, we developed a Large Language Model (LLM) assistant to support the search and filtering of documents.\\
\textbf{Aims}: To develop and validate an LLM-based tool that can support researchers in performing the search and filtering of documents for an MVLR without compromising the rigor of the research protocol. \\
\textbf{Method:} We applied sound engineering practices to develop an on-premises LLM-based tool incorporating Retrieval Augmented Generation (RAG) to process candidate sources. Progress towards the aim was quantified using the Positive Percent Agreement (PPA) as the primary metric to ensure the performance of the LLM-based tool. Convenience sampling, supported by human judgment and statistical sampling, were used to verify and validate the tool’s quality-in-use.\\
\textbf{Results:} The tool currently demonstrates a PPA agreement with human researchers of \textbf{\GetValue{FinalPPAITeration2}} for sources that are not relevant to the study. Development details are shared to support domain-specific adaptation of the tool.\\
\textbf{Conclusions:} Using LLM-based tools to support academic researchers in rigorous MVLR is feasible. These tools can free valuable time for higher-level, abstract tasks. However, researcher participation remains essential to ensure that the tool supports thorough research.
}




\maketitle


\section{Introduction}\label{Sec_Intro}
Multi-Vocal Literature Reviews (MVLRs) are valuable research methods for gathering evidence from the state of practice in software engineering \cite{garousi_MVLR_2016}. The method has become an increasingly popular way to tackle industry practices in the empirical software engineering research community. From a research design perspective, when performing MVLRs, the sources the researcher is working with are unstructured and vary in the depth of technical details. Therefore, the researcher typically works with materials that have sparse information density. This is time-consuming, as a significant amount of effort must be invested in understanding and evaluating sources that will not be considered during the analysis and reporting phase of the study. Despite this time-consuming process, MVLRs remain a relevant research method, particularly for domains where the desired evidence isn't readily available in peer-reviewed academic literature. The inherent trade-off is justified by the unique value of the information uncovered, which can significantly advance the research domain.

Our research team has been developing theory and evidence on the challenges of testing Context-Aware Software Systems (CASS), with a high interest in researching how these systems are being tested in the industry. Consequently, we designed and conducted a Multi-Vocal Literature Review (MVLR) to explore how the automotive industry was testing CASS  \cite{DreamTeam_TII_2025}. Within these MVLR in the Automotive domain, the effort/value tradeoff issue of MVLR was evident yet manageable. However, when the team began instantiating the research protocol for the avionics industry, it encountered a domain with significantly more variation in sources, which presented an industry with a complex supply chain and different industry sub-sectors. The researchers' result was a substantial increase in potential sources of interest by several orders of magnitude compared to the automotive industry (see Section \ref{sec:BackgroundResearch}). 

Therefore, we decided to explore the application of Large Language Models (LLMs) to support the screening and selection of sources.
This paper presents the design, development, and validation of an on-premises LLM-based tool to support researchers in searching and filtering sources that present unstructured information. As mentioned, this is typical of MVLR studies, where researchers deal with sources that lack a structured internal organization. 
To advance our research agenda, we aimed to produce a tool that would support the researchers in the MVLR without compromising the rigor of the MVLR protocol. As such, we took special care that:
\begin{itemize}
    \item The LLM adheres to the criteria outlined in the research protocol. This means that the tool closely follows the inclusion and exclusion criteria defined in the protocol, and care was taken to minimize the risk of hallucinations; 
    \item The precision and recall risks associated with the tool are comparable to the risks introduced by human researchers. This means that our starting point is that even for highly cohesive teams, there is a certain degree of variation in how humans interpret selection criteria. Our engineering process utilized measures of statistical agreement to ensure that the LLM tool's implementation of the inclusion/exclusion criteria was comparable to how the research team understood them. In short, the bias introduced by the tool should be similar to the bias introduced by any member of our research team.
\end{itemize}
The result is an on-premises LLM-based tool that can screen and process multiple sources, determining whether human researchers should invest time in researching them. The effectiveness of the tool in discarding sources is comparable to that of human researchers, as calculated through the Positive Percent Agreement (PPA) between humans and the LLM-based tool of \textbf{\GetValue{FinalPPAITeration2}}. 

This paper contributes to the research body of LLM to support academic original research by providing a reusable tool for other MVLRs. This tool frees valuable researcher time that can be made available for higher-level activities. Furthermore, the  development process followed open science principles, ensuring reproducibility. Supporting materials include all artifacts of the engineering process (prompts, code, data, and analysis) for other teams to replicate our results in their research domains of interest. The tool is based on open-source LLMs and can be run on-premises, without relying on proprietary LLMs or commercial API providers.

This paper is organized as follows: section \ref{sec:BackgroundResearch}, frames the motivation for developing the tool in light of the research agenda and its challenges. Section \ref{sec:RelatedResearch}, presents background research of the applications of LLM to support different aspects of secondary studies, including MVLRs. Section \ref{sec:llm_based_tool_design}, details the design and development of the LLM-based tool. Section \ref{sec:results}, presents the empirical results that show how the tool can be used to filter sources for our MVLR. Section \ref{sec:Discussion} presents some interesting points that have to be taken into consideration for researchers trying to adapt the tool to their research domain (including a discussion of threats to validity). Finally, in the section \ref{sec:Conclusion} present our conclusions and future work.

\section{Problem Characterization and Motivation}\label{sec:BackgroundResearch}

This section aims to provide a brief context for the research agenda that led to the development of the LLM-based tool.  As we mentioned in the introduction, our research interest lies in advancing the understanding and developing theories to support the Testing of CASS. This research line has been ongoing since 2015, \cite{Matalonga2017, Santos2017}. 
As a result, a significant portion of this team's research efforts has been dedicated to identifying approaches in peer-reviewed literature that align with this perspective. Needless to say, we do not claim that these are the only systematic reviews concerning the testing of CASS. Other groups have conducted secondary studies on the topic \cite{siqueira_testing_2021}. However, thus far, our research has uncovered very limited evidence of testing approaches that encompass all these criteria \cite{matalonga_alternatives_2022}.

When examining the scientific peer-reviewed literature, a key issue with the limited number of sources covered by these secondary studies is their restricted capacity for generalization. These works argue that the core problem stems from the lack of standardization in the terminology of our discipline. For instance, terms like "testing," which one might expect to have a clear and consistent definition, are subject to varying interpretations. For example, the ISO 29119 \cite{ISO_29119} defines reviews as "static testing," whereas the ISTQB \cite{Thompson2024_ISTQB} defines testing as the dynamic execution of a test item. Consequently, researchers must exercise significant judgment when selecting relevant works. 
A related challenge involves the interpretation of the term "context." It is an overloaded term with numerous—albeit correct—applications that do not pertain to the CASS domain. Combined with the limitations of keyword-based search engines, this term overloading affects the precision and recall ratios of secondary studies. As a result, it is common to observe secondary studies that screen numerous sources, only to select a relatively small number of them. When performing MVLR, the problems of the nomenclator remain, but the unstructured nature of non-peer-reviewed information exacerbates them. Not only do researchers have to exercise judgment to filter sources relevant to their review, but they also have to do so with sources that contain very little information density.

This is a necessary trade-off that researchers accept to advance our research agenda. In the following subsection, we aim to demonstrate that when transitioning from the Automotive industry to the avionics industry, the effort required to overcome information challenges would have effectively halted the advancement of research.

\subsection{An instance of the MVLR Research Protocol for the Automotive Domain} 
In this section, we present the key sections of the Automotive domain MVLR protocol that are relevant to understanding the need for the LLM-based tool when compared with the instance for the Avionics domain.

\textbf{Aim:} By conducting these MVLRs, we aim to \textit{uncover evidence on how \textbf{industries deploying CASS} report their work with the dynamic testing process regarding such software systems.} As a result, the first step was to identify these industries. Of those industries, we started with the Automotive Industry \cite{DreamTeam_TII_2025}. The full protocol is available in \cite{DT_Automotive_Protocol_2023}.

\textbf{Searching for sources}. In searching, we utilized a catalog of companies within the automotive industry. This catalog was initially populated based on our existing knowledge and further enhanced using Search String I (see Table \ref{Table_AutomotiveSearchString}). Sources were subsequently identified through the application of Search String II, with the results contributing to an iterative refinement of the company catalog. Notably, the search was intentionally restricted to PDF documents. This decision was made because PDFs published by these companies will likely contain curated and deliberate information that the companies wish to share publicly.

\begin{table}[h]
\caption{Search Strings used in the Automotive instance of the MVLR protocol}\label{Table_AutomotiveSearchString}%
\begin{tabular}{@{}llll@{}}
\toprule
Search String I & Search String II \\
\midrule
software test* AND
automotive filetype:pdf    & software test* AND
[company] filetype:pdf  \\
\botrule
\end{tabular}
\end{table}

\textbf{Inclusion criteria}: The following three inclusion criteria were developed for this protocol.

\begin{enumerate}[label=IC\arabic*]
    \item The documents must be published in PDF format.
    \item The document describes how the software system:
    \begin{enumerate}[label=\alph*)]
        \item is tested (example reports), or
        \item can be tested (future proposal), or
        \item should be tested (standards).
    \end{enumerate}
    \item The document focuses on testing an automotive software system fitting the academic definition of CASS.
\end{enumerate}

It is essential to note that, for the Automotive instance, the search was conducted manually using the Google Search Engine, and the filtering was performed through agreements, where each source was reviewed by three researchers, who then voted on whether to include it in the study.   We reviewed the first 100 Google search results for each company, and accommodated Google's algorithm variation by searching with all our users and then merging the results. 
This entailed manually screening almost 1800 Google search result titles, which led to the identification of 120 potential PDFs. The subsequent voting process narrowed this list to 36, which were then processed for data extraction and analysis. 
As it is common with MVLR, the sources varied widely in shape, size, and format. This inherent variability presents a significant challenge for researchers tasked with identifying, filtering, and consolidating evidence. \footnote{Our attempt at categorizing the PDF types in the sample revealed a spectrum from short advertising documents to extensive master's theses (see \cite{DreamTeam_TII_2025}).}

While 120 sources remained a feasible undertaking, researchers with experience in executing MVLR will recognize the substantial redundancy inherent in this process. Multiple researchers review the sources to mitigate bias, and significant time is allocated to voting and discussion to achieve agreement in inclusion/exclusion decisions.\footnote{The final number of studies included in \cite{DreamTeam_TII_2025} was 20. As sources are rejected during data extraction too.} 

\subsection{Challenges with the Avionics Industry.}

The situation became much more complex when we focused on the avionics domain. First, the number of companies in the avionics domain is vastly bigger. The industry is divided into large, complex supply chains that are far larger than the main known brands. We made the scoping decision only to include those companies for which a \textit{non-military manned aircraft will fly and be branded after a company name}. This resulted in 175 companies (for comparison, the automotive search included 18 companies). This lead to our first design decision.

\textit{Automate the search process through the Google API}. Initial attempts to automate this process revealed that our population of potential documents would yield approximately 8,000-10,000 results. These possible sources would share the same complexities for the automotive domain:

\begin{itemize}
    \item Sparse information density
    \item Most sources would be rejected; yet, the team should still review them.
\end{itemize}

\begin{figure}[b]
\centering
\includegraphics[width=\textwidth]{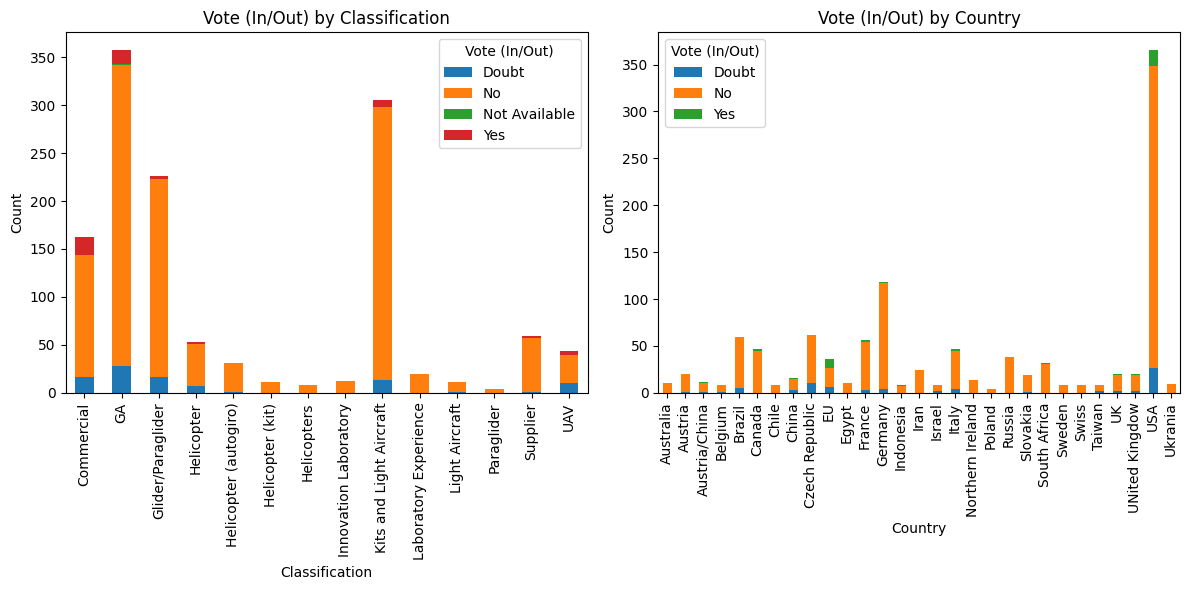}
\caption{Initial assessment of potential sources for the avionics domain MVLR. A) shows potential sources grouped by a custom classification. B) shows potential sources grouped by country of origin of the related company}\label{Fig_DistributionOfInitialScreening}
\end{figure}

To confirm these hypotheses, we explored several sampling approaches. Figure \ref{Fig_DistributionOfInitialScreening} illustrates the outcomes of one such approach, where we tried to identify sampling strategies by providing a bespoke classification of the sources. The overall conclusion was that, regardless of the stratification of the sources, the majority of the material would not be relevant for the research goals. Therefore, identifying the potential sources would be effort-intensive \footnote{We documented all sampling attempts in the Jupyter notebook 'SamplingExcercise', available in our replication package: \href{https://git-lab.cos.ufrj.br/contextaware/llm}{https://git-lab.cos.ufrj.br/contextaware/llm}}.


\begin{table}[h]
\centering
\caption{Indicative results of Screening early Search results for the avionics industry}\label{Table_ScreeneingResults}
\begin{tabular}{lllll}
\toprule
Population & Sampled & Include & Doubt & No \\ \midrule
8482       & 1506    & 60      & 108    & 1338 \\ \botrule
\end{tabular}
\end{table}

These initial results highlight the tradeoff between the perception of time required to filter the sources and the benefits gained from our research. Our experience with MVLR biases our perception of time to be proportional to the number of sources, and also multiplied by the number of researchers assigned to each source, where in the past we have worked with three researchers per source to work out voting and disagreements. The initial screening found that less than 4\% (60/1,506) of the results were potentially relevant. And even so, the quality of the material in those documents was dubious, indicating that the final sample would likely be made up of an even smaller percentage. Therefore, without a different approach that could automate part of the screening and filtering task, the research line would have essentially come to a halt. Consequently, we turned our attention to utilizing Large Language Models to review and select (or, as the section \ref{sec:llm_based_tool_design} will show, discard) the sources so that our attention and effort will be dedicated to reviewing sources that would be much more likely to have relevant information for our research.

\section{Related works on the applications of LLM to support research and higher-abstraction tasks}\label{sec:RelatedResearch}
Recent advancements in deep learning and natural language processing (NLP) have sparked increased interest in automating Systematic Literature Reviews (SLRs). Large Language Models (LLMs), particularly general-purpose models such as ChatGPT, have been studied across various disciplines—including medicine, healthcare, education, and the social sciences—for their potential to assist at different stages of the SLR process \cite{waseem2023}. Prior research has reported the use of LLMs to support tasks such as formulating research questions, refining Boolean search strings, screening titles and abstracts, applying inclusion and exclusion criteria, extracting data from primary studies, and synthesizing evidence. For example, Alshami et al.~\cite{alshami2023} and Wang et al.~\cite{wang2023} investigated LLMs' ability to construct effective search queries, while others~\cite{guo2024, khraisha2024, robinson2023, wilkins2023} focused on their performance in automating the study selection process. Further contributions, like Gupta et al. \cite{gupta2023}, explored their use in generating novel review ideas and synthesizing insights from selected studies. 

Within Software Engineering (SE), empirical research on this topic is comparatively limited but growing. Alchokr et al.~\cite{alchokr2022} examined the application of deep-learning-based language models to support SLRs in SE, highlighting their ability to cluster and filter relevant studies and reduce reviewer workload. Watanabe et al.~\cite{watanabe2020} proposed a supervised text classification approach to facilitate updating existing reviews, offering early empirical evidence of automation benefits in SE-specific study selection tasks.

In addition to LLM-focused work, other studies have explored AI-based support for SLR-related tasks in software engineering (SE) through various lenses. Girardi et al.~\cite{girardi2023}, for instance, conducted a systematic review of deep learning methods for software defect prediction, indirectly highlighting the integration of data-driven techniques into empirical SE research pipelines. Similarly, Necula et al.~\cite{necula2024} presented a systematic mapping study on applying NLP techniques in requirements engineering, demonstrating the broader applicability of language models across SE subdomains. Al-Shalif et al.~\cite{alshalif2024} further reviewed metaheuristic feature selection strategies for text classification, providing insights relevant to automating inclusion/exclusion decisions.

Despite promising results, several studies concur that LLMs should be utilized as complementary tools, rather than replacements for human decision-making. Concerns include limited transparency, hallucinated content, reliance on potentially outdated training data, and lack of access to closed-access sources~\cite{anghelescu2023, mahuli2023, najafali2023, qureshi2023}. Huotala et al.~\cite{huotala2024}, for instance, found that LLM-based abstract screening did not outperform human reviewers in SE. Felizardo et al.~\cite{felizardo2024} offered a more comprehensive evaluation, applying ChatGPT-4 to replicate the study selection phase of two SLRs in SE. While the model achieved encouraging accuracy levels (75.3\% and 86.1\%), the authors emphasize the risks of false exclusions and sensitivity to prompt design, recommending its use primarily as a secondary reviewer or support mechanism for novice researchers.

In the context of LLMs in MVLR, Khraisha et al. \cite{khraisha2024} performed a feasibility study to evaluate GPT-4's performance in screening and extracting data for secondary studies. While their results indicated significant variation in the tool's effectiveness, they highlighted the importance of tailoring prompts to each specific task.

Despite this growing body of work, the use of LLMs in the context of \textit{Multivocal Literature Reviews} (MLRs) or \textit{Grey Literature Reviews} (GLRs) remains largely unexplored. Most empirical evaluations to date have focused on structured, peer-reviewed sources, with limited attention given to heterogeneous or informal content types that are typical of grey literature. In summary, while preliminary evidence supports the usefulness of LLMs in specific tasks related to literature review, their systematic integration into multivocal review processes remains an open area for research. In particular, we have observed that most research aims to evaluate LLMs' capacity to perform tasks, using datasets from previous (and published secondary studies). To the best of our knowledge, this is the first tool specifically engineered to support an ongoing research project.


\subsection{LLM prompt best practices}\label{Sec_PETheory}
Large Language Models (LLMs) hold significant potential to create high-quality digital content. However, LLMs frequently present challenges in controlling the quality of the responses \cite{dang2022_ArXiv_Prompts}. The practice of prompting generative models allows end-users to creatively assign novel tasks to LLM-based systems in an ad hoc manner, merely by articulating them. Nonetheless, for most end-users, formulating effective prompts remains predominantly reliant on trial and error \cite{dang2022_ArXiv_Prompts}.

Thus, prompt engineering has become an important technique for enhancing the capabilities of pre-trained large-scale language models. This approach involves the strategic formulation of task-specific instructions, known as prompts, to guide the models' generation of outputs without requiring modifications to their internal parameters. The relevance of this technique is particularly evident due to its transformative impact on the adaptability and functional versatility of LLMs across different domains and application contexts \cite{sahoo2025_ARXiv_SSPE}.

Many prompt engineering approaches have been proposed to improve interaction with LLMs or to obtain better results \cite{Liu_2023_PromptingMethods, sahoo2025_ARXiv_SSPE, Marvin_2023_Prompt}, such as: (i) focusing on the amount of response examples included in the prompts between zero-shot to few-shot prompting; (ii) according to the form of reasoning and logic of the prompts, such as chain-of-thought (CoT), self-consistency, logical CoT; (iii) varying in the approach to reduce hallucinations such as react prompting, chain-of-verification or retrieval-augmented generation (RAG); (iv) in the way of generating programming code such as scratchpad prompting or program-of-thoughts (PoT) prompting; (v) for personalizing content generation such as expert prompting, role-play prompting, among others \cite{xu2025_expertprompting, Liu_2023_PromptingMethods, Chen_2023_Unleashing_prompt, Minjun_2025_prompt}. Therefore, selecting the best prompt technique depends on several factors: user intent, model understanding, domain vocabulary, clarity and specificity, and constraints such as response size or expected response format \cite{Sabit_2023_PE4ChatGpt}.

Even when choosing the correct prompt technique for a specific case, the best LLM responses are often not obtained on the first attempt, requiring effort to refine the text submitted to the models to achieve better results. Thus, prompts need to be refined in several ways, through pure experimentation, reusing a library of prompts, reverse engineering, using the models themselves to optimize the prompts, trying to reduce the computational cost of processing the response, or decomposing a complex prompt and chaining it into a sequence of subtasks \cite{Clariso_Cabot_2023, Marvin_2023_Prompt}.

However, applying good prompting techniques and refining the quality of prompts may often not be enough to prevent LLMs from producing so-called hallucinations, which consist of generating content that appears factual to the reader but is ungrounded \cite{Tonmoy_2024comprehensive_allucinations}. This hallucination potential is inherent to LLMs, which are exposed to massive amounts of text data during training, allowing them to achieve impressive linguistic fluency and also to extrapolate information from the training data \cite{Tonmoy_2024comprehensive_allucinations}. An alternative within prompt engineering to mitigate hallucinations is the Retrieval-Augmented Generation (RAG) approach \cite{Lewis_2020_RAG}. RAG enhances LLM-generated text by combining the generation of the answer with an earlier retrieval step, in which external expert knowledge is used to supplement the information. This integration improves the accuracy of information retrieval and enables the model to provide more precise answers to queries involving up-to-date information or very specific facts \cite{Minjun_2025_prompt}. 

Given its ability to improve filtering, classification, synthesis, and extraction of text content, prompt engineering approaches have also been used to support systematic literature review activities. Zero-shot prompting strategies have been used to assist in formulating research questions and creating initial eligibility criteria \cite{mahuli2023}, in developing search strategies  \cite{qureshi2023}, in screening titles and abstracts \cite{guo2024, huotala2024} and in data extraction \cite{mahuli2023}. One-shot and few-shot prompting are also used to screen titles and abstracts, providing "seed papers" for the LLM \cite{alchokr2022, huotala2024}. In this sense, Chain-of-Thought prompting approaches have also been applied in screening titles and abstracts, where models are encouraged to decompose systematic literature review activities into smaller steps \cite{huotala2024}. Role-play Prompting/Expert Prompting have also been employed, for example, to instruct the LLM to assume a specific role in systematic literature reviews, such as that of a “researcher who screens titles of scientific articles” or a domain “expert” \cite{guo2024}.


\section{The LLM-Based Tool for Searching and Filtering in MVLRs} \label{sec:llm_based_tool_design}

This section presents the development process of the on-premises LLM-based tool that we specifically designed, built, and validated to support the execution of a Multivocal Literature Review (MVLR). The tool was developed to automate two central and effort-intensive phases of the MVLR protocol: (i) the identification of potentially relevant documents from grey literature and (ii) their preliminary screening based on the inclusion and exclusion criteria defined in our research protocol. The ultimate goal was to enable scalability without compromising methodological rigor, ensuring that the automated decisions remain consistent with those made by human researchers.


To this end, we defined and automated a structured process for the review. This process was designed to reflect our MVLR protocol, in order to make sure that the results obtained from the tool would be useful for continuing our research. Details concerning the design assumptions and implementation strategies are discussed in the following subsections.

\subsection{LLM-Based Tool Design}

\begin{figure}
    \centering
    \includegraphics[width=0.8\linewidth]{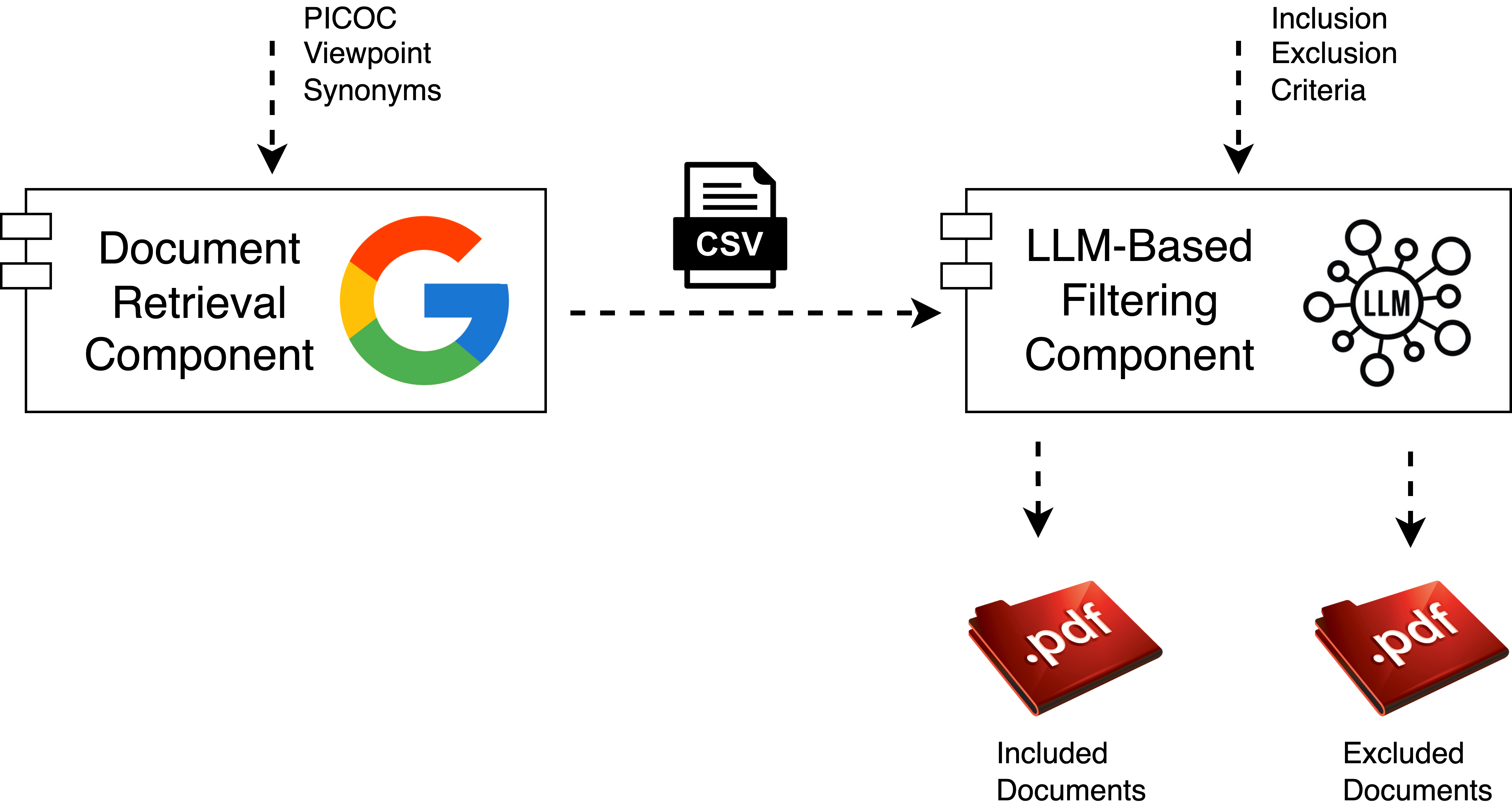}
    \caption{The LLM-Based Tool Pipeline Design}
    \label{fig:tool-architecture}
\end{figure}

The tool supports two primary activities within an MVLR: 1) a document retrieval phase, guided by domain-specific search strings executed via web search engines, and 2) a filtering phase, where each document is evaluated against protocol-driven criteria. As illustrated in Figure~\ref{fig:tool-architecture}, the tool implements these activities using a pipeline structure composed of two main components, each corresponding to one of these activities.

The first component, the \textit{Document Retrieval Component}, receives as input one or more structured search strings, typically derived from a PICOC formulation. It uses Google as the underlying search engine to retrieve publicly available documents in PDF format, which are then passed to the next stage of the process.

The second component, the \textit{LLM-Based Filtering Component}, takes as input the documents retrieved in the previous phase and evaluates each of them individually. A Retrieval-Augmented Generation component (RAG) is employed to process each document and incorporate the contents into the knowledge of the LLM model. The goal is to determine whether the contents of the document satisfy the inclusion criteria and do not meet any exclusion criteria, as defined in the MVLR protocol. This decision process is delegated to a large language model, which evaluates the pertinence of the content of the documents to the research.

In addition to the two core functional requirements, the key measure of success was that the tool should not introduce more bias than what is typically associated with human reviewers applying the same protocol. These requirements were assessed during the Prompt Engineering phase of the process and are explained in section \ref{sec:results}. Furthermore, we added the restriction that the tool had to be executable on locally available and resource-limited hardware, without relying on commercial cloud services. It also had to be compatible with open-source technologies to ensure transparency, replicability, and full adherence to the principles of open science. \footnote{Source code, datasets with execution results and Jupyter notebooks used for data analysis are available in: \href{https://git-lab.cos.ufrj.br/contextaware/llm}{https://git-lab.cos.ufrj.br/contextaware/llm}}

\subsection{LLM-Based Tool Implementation}

The implementation of the LLM-based tool reflects the structure and objectives defined during the design phase and was instantiated in the context of a Multivocal Literature Review focused on the avionics domain. The protocol was defined to investigate \textit{how context-aware software systems are tested in the civil aviation industry}, and it included explicit inclusion and exclusion criteria, as well as domain-specific terminology derived from a PICOC-based formulation. These elements guided the configuration of both tool components.

The implementation consists of two independent scripts. The first automates the retrieval of grey literature documents from the web, while the second filters the retrieved documents by assessing their relevance concerning the MVLR protocol. The following subsections describe the implementation details of each component.

\subsubsection{Document Retrieval Component Implementation}

The automated search component of the tool was designed to retrieve grey literature documents in PDF format by querying the Google Custom Search API. Queries were constructed to reflect two specific dimensions of the PICOC framework: \textit{Population} and \textit{Intervention}, instantiated through domain-specific controlled vocabularies selected according to the scoping decisions defined in the MVLR protocol.

The sets of terms used for query formulation are reported below:

\begin{itemize}
    \item \textbf{Population terms:} \texttt{Aviation}, \texttt{Aeronautics}, \texttt{Aerospace}, \texttt{Flight Science}, \texttt{Air Transportation}, \texttt{Aeromechanics}, \texttt{Air Navigation}, \texttt{Avionics}, \texttt{Airspace Management}.
    \item \textbf{Intervention terms:} \texttt{testing}, \texttt{test}, \texttt{verification}, \texttt{validation}, \texttt{quality assessment}.
\end{itemize}

To support the search within the avionics domain, we developed a Java-based tool that automates the retrieval and download of PDF documents using the Google Custom Search API. The tool executes structured queries based on predefined combinations of domain-specific and testing-related keywords.

It accepts two sets of search terms—one for domain concepts (e.g., \textit{Aviation}, \textit{Aeronautics}) and one for software testing activities (e.g., \textit{testing}, \textit{validation})—and supports two query strategies: (i) a basic strategy combining a single term from each category using \texttt{AND}, and (ii) a broader \texttt{OR}-based strategy that merges all domain terms with a single testing-related keyword. The final query string includes the \texttt{intext:} directive and restricts results to PDF format using \texttt{filetype:pdf}.

Once the query is built, the tool interacts with the Google API to iteratively retrieve result pages in batches of ten, extracting candidate URLs from the JSON response. The overall retrieval procedure is described in detail in Algorithm~\ref{alg:retrieval}.

To comply with API constraints such as rate limits and pagination, the tool increments the \texttt{start} index to access subsequent result pages. It also validates HTTP responses, filters by MIME type (\texttt{application/pdf}), and handles download failures gracefully, accounting for invalid links, redirects, or paywalled content.

Since the Google Custom Search API limits each query to a maximum of 100 results, we adopted a structured formulation strategy to increase coverage. For each intervention term, we generated a Boolean query that combines all population terms using the OR operator and joins the result with the intervention term using AND. This led to five distinct queries, each executed independently to retrieve up to 100 PDF results. This approach maximized recall within the API constraints while preserving thematic coherence across queries.

The retrieval process enabled scalable and reproducible collection of domain-specific grey literature, resulting in a corpus of technical and industrial documents on software testing in aeronautics. This dataset was used in the subsequent phases of the multivocal review.

\begin{algorithm}[h]
\caption{Automated Retrieval of Domain-Specific PDFs links}
\label{alg:retrieval}
\begin{algorithmic}[1]
\State \textbf{Input:} 
\begin{itemize}
  \item \texttt{populationTerms} $\gets$ \{"Aviation", "Aeronautics", "Aerospace", "Flight Science", "Air Transportation", "Aeromechanics", "Air Navigation", "Avionics", "Airspace Management"\}
  \item \texttt{interventionTerms} $\gets$ \{"testing", "test", "verification", "validation", "quality assessment"\}
\end{itemize}
\State \textbf{Output:} Downloadable PDFs links in .CSV and search logs

\ForAll{\texttt{intervention} in \texttt{interventionTerms}}
    \State Construct query: 
    \State \hspace{1em} \texttt{(intext:"term$_1$" OR ... OR intext:"term$_n$") AND intext:"intervention" AND intext:"software" filetype:pdf}
    \State Create output folder and initialize TXT and CSV log files
    \State \texttt{startIndex} $\gets$ 1
    \While{more results are available}
        \State Compose API URL with \texttt{startIndex} and query
        \State Send HTTP GET request and parse JSON response
        \ForAll{items in response}
            \State Extract \texttt{link} and write to TXT log
            \If{\texttt{link} ends with `.pdf''}
                \State Validate content type and HTTP response
                \If{valid}
                    \State Log success in CSV
                \Else
                    \State Log failure in CSV
                \EndIf
            \EndIf
        \EndFor
        \State \texttt{startIndex} $\gets$ \texttt{startIndex} + 10
    \EndWhile
\EndFor
\end{algorithmic}
\end{algorithm}

\subsubsection{LLM-Based Filtering Component Implementation}

The filtering component of the tool was designed to assess the relevance of documents retrieved from the automated search phase, based on a set of predefined inclusion and exclusion criteria.

The relevance of each document was assessed based on a set of predefined inclusion and exclusion criteria. These criteria reflected our scoping decisions and were iteratively refined during the prompt engineering process to align with the objectives of the MVLR.

Documents were considered for inclusion if they met one or more of the following conditions:

\begin{enumerate}[label=IC\arabic*]
    \item The document concerns an aircraft that is manned or piloted;
    \item The aircraft operates within the civil aviation domain;
    \item The document indicates the existence of digital components or software in the aircraft;
    \item The document describes the design, execution, or reporting of the testing of aircraft systems;
    \item The document describes software testing techniques, technologies, processes, or standards, and;
    \item The described context is applied in the industry.
\end{enumerate}

Conversely, documents were excluded if they met any of the following conditions:

\begin{enumerate}[label=EC\arabic*]
    \item The document is an operating or installation manual;
    \item The content focuses on military applications;
    \item The subject is related to spacecraft;
    \item The document only describes static analysis techniques, or;
    \item The document refers exclusively to the design or execution of hardware-only avionics systems.
\end{enumerate}

Together, these criteria expressed our scoping decisions regarding the study. The wording of the inclusion and exclusion criteria underwent multiple iterations during the prompt engineering phase (see section \ref{sec:validation_in_use}), as we refined how to present them to the LLM.


Algorithm \ref{Alg:Filtering} presents a high-level overview of the process. Essentially, the process has three stages. Access and download the content, then encode it. Finally, submit the content along with the question and prompt to the LLM for evaluation. The process received three inputs: the list links to the PDF documents (in CSV format) that are the results of the execution of the \textit{Document Retrieval Component}, the \textit{prompt} , which provides the context of the research, including the inclusion and exclusion criteria, and the \textit{question} that offers the intervention for the LLM. Both \textit{prompt} and \textit{question} iteratively evolved as we fine-tune the behavior of the LLM-based filtering components (see section \ref{sec:validation_in_use}).

\begin{algorithm}[h]
\caption{Algorithm for the LLM-Based component}
\label{Alg:Filtering}
\begin{algorithmic}[1]
\State \textbf{Input:} \\
\begin{enumerate}
    \item CSV from search script with the List of Documents, format (ID, URL) 
    \item Prompt \#Provides the context, IC and EC
    \item Question \# Provides the expectation
\end{enumerate}
\State \textbf{Output:} 
\begin{enumerate}
    \item A set of Documents Downloaded and evaluated
    \item An evaluation log with a decision for each document
\end{enumerate}
\ForAll{Document doc in the CSV of documents}
    \State Access doc (wait a return for 12 seconds)
    \State Download the document
    \If{document is downloaded}
        \State Encode()
            \State Convert the PDF document to text and normalize it.
            \State Collect relevant information from the embedded content.
        \State Filter()
            \State Send \textbf{Prompt}, \textbf{Question} , and encoding to the LLM for it to make an evaluation about the pertinence of the Document.
            \If{answer is \textbf{YES} or \textbf{DOUBT}}
                \State Save the document in the folder \textbf{PDF}
            \Else
                \State Discard the document
            \EndIf
            \State Register the answer in the evaluation log
    \Else
        \State Register NOT AVAILABLE in the evaluation log
    \EndIf
    \State reset environment
\EndFor
\end{algorithmic}
\end{algorithm}

The implementation of this algorithm in Python can be quite straightforward. However, communicating with the LLMs has two key points, which we describe in the following subsections.

\paragraph{RAG and encoding}\label{sec:RAG_Encoding}
In the context of RAG, the document must be prepared so that the LLM can process it. As mentioned, we sought out to use open-source technologies that could be executed within the constraints of our computing hardware. 


Our selection of \textit{Llama} and \textit{mxbai-embed-large}, both open-source, eliminates licensing constraints that might otherwise impede the publication or sharing of our methodology. Furthermore, all computing is executed locally, preventing any data transmission to third parties, and showing that LLM can contribute to science with relatively low-end hardware. It also conveys the reproducibility of our work, allowing other researchers to replicate, build upon, and extend our findings easily.

PDF content is extracted and structured into plain text without reliance on external services. Algorithm~\ref{alg:pdf_chunking} outlines this stage of the RAG pipeline, where document data is parsed and prepared for embedding. The resulting text, together with its embeddings, feeds into the LLM's inference stage. One key insight from this process was to introduce overlap during sentence chunking (Lines~16–23), enhancing contextual continuity across chunks and improving inference quality.
To meet hardware constraints and adhere to open science principles, we selected an open-source foundational model that could run on available machines, avoiding reliance on proprietary cloud-based services such as OpenAI ChatGPT.
These hardware constraints (see section \ref{sec:validation_in_use} for the hardware specifications) determined the maximum model size that could be employed for the filtering task. We selected two models: one for generating embeddings (\texttt{mxbai-embed-large}\footnote{\url{https://ollama.com/library/mxbai-embed-large}}) and one for assessing the relevance of sources (\texttt{dolphin-llama3}\footnote{\url{https://ollama.com/library/dolphin-llama3}}).

\begin{algorithm}[h]
\caption{PDF Text Extraction and Chunking}
\label{alg:pdf_chunking}
\begin{algorithmic}[1]
\Require \texttt{pdf\_name} \Comment{Filename of the PDF to process}
\Ensure Chunked text is written to \texttt{vault.txt}

\State \texttt{file\_path} $\gets$ \texttt{pdf\_name}
\If{\texttt{file\_path} is valid}
    \State Open \texttt{pdf\_file} from \texttt{file\_path}
    \State \texttt{pdf\_reader} $\gets$ PyPDF2 reader of \texttt{pdf\_file}
    \State \texttt{text} $\gets$ empty string

    \ForAll{page in \texttt{pdf\_reader.pages}}
        \State \texttt{extracted} $\gets$ page.extract\_text()
        \If{\texttt{extracted} is not empty}
            \State Append \texttt{extracted} to \texttt{text}
        \EndIf
    \EndFor

    \State Normalize whitespace in \texttt{text}
    \State Split \texttt{text} into \texttt{sentences} at punctuation boundaries
    \State \texttt{chunks} $\gets$ empty list
    \State \texttt{current\_chunk} $\gets$ empty string

    \ForAll{sentence in \texttt{sentences}}
        \If{length(\texttt{current\_chunk} + \texttt{sentence}) < 1000}
            \State Append \texttt{sentence} to \texttt{current\_chunk}
        \Else
            \State Append \texttt{current\_chunk} to \texttt{chunks}
            \State \texttt{current\_chunk} $\gets$ \texttt{sentence}
        \EndIf
    \EndFor

    \If{\texttt{current\_chunk} is not empty}
        \State Append \texttt{current\_chunk} to \texttt{chunks}
    \EndIf

    \State Write each \texttt{chunk} as a line in \texttt{vault.txt}
\EndIf
\end{algorithmic}
\end{algorithm}

\paragraph{Prompt Engineering}

Our prompting strategy encompasses various techniques, including Role-Play and Expert prompting (where we define a role to be assumed), Few-Shot prompting (where we include examples of expected outputs), and Chain-of-Thought prompting (where we explicitly define the rules and reasoning steps to be followed). This strategy allowed for mapping the inclusion and exclusion criteria, as described in the research protocol, to the instructions represented in the prompt. However, the tailoring required a series of interactive trials to ensure the prompt and question were adequate to support the filtering.  This \textit{prompt engineering} process ran from November 2024 to February 2025.


Therefore, it is possible to observe five prompt and question versions, which evolved based on the indications of suitability provided by the collected measures. The evolution of the questions (see table \ref{tab:UserQuestionsEvo}) and prompts was due to the level of detail and the combination of inclusion and exclusion criteria used to influence the LLM's behavior. 

\begin{table}[h]
    \centering
    \begin{tabularx}{\textwidth}{lX}
        User Questions & Description \\
        \hline
        UQ0 & Context-Aware Software Testing \\
        UQ1 & Does the document regard the testing of context-aware software systems?\\
        UQ2 & Would you select this document to support your activity of avionics context-aware software systems? \\
        UQ3 & Is this document relevant and suitable for supporting the testing of avionics context-aware software systems?\\
        UQ4 & Would you choose this document to support testing context-aware avionics software systems in the industry?\\
        \hline
    \end{tabularx}
    \caption{versions of the User Question}
    \label{tab:UserQuestionsEvo}
\end{table}

As it happened with the Questions, the single-shot prompt also evolved during the \textit{prompt engineering} process. Table \ref{tab:PromptEvo} presents a mapping between the questions, the versions of the prompt, and the platform, and samples with which the results were evaluated. After each modification of the prompt or questions, we would execute the Algorithm against a suitable sample (see section \ref{sec:validation_in_use} for the details of the evaluation process and evolution of the PPA metric).

For exemplification, we present the differences between V0.0 (see table \ref{tab:PromptV0.0}) and V4.1 (see table \ref{tab:Prompt V4.1}) of the prompt\footnote{Major versions of the prompts used are available in the replication package \href{https://git-lab.cos.ufrj.br/contextaware/llm}{https://git-lab.cos.ufrj.br/contextaware/llm} together with comparison of the main changes among each major version} . The evolution of the prompt enabled the submission of a detailed set of instructions to the LLM.   

We conducted an exploratory analysis to assess the influence of the temperature parameter. All versions were initially executed with a temperature of 0.5. In subsequent runs involving versions 3 and 4, the temperature was lowered to 0.1. The results indicated that this parameter had a limited impact, introducing a minor trade-off: a marginal improvement in response confidentiality for a slight increase in processing time.

\begin{table}
    \centering
    \begin{tabularx}{\textwidth}{|X|}
    \hline
     You are a software tester specialized in selecting documents talking about testing context-aware software systems of aircraft. You always select a document when all of these five rules are satisfied: \\
     1 - An aircraft manned or piloted;\\2 - An aircraft operating within civil aviation;\\3 - The document indicates the existence of digital components or software in the aircraft;\\4 - The document describes the design, execution, or reporting of the testing of aircraft systems;\\5 - The document describes software testing techniques, software testing technologies, software testing processes, or software testing standards. \\\\You always reject a document when any of these four rules are satisfied:\\1 - The document is an Operating or installation manual;\\2 - The document describes Military applications;\\3 - The document describes Spacecraft;\\4 - The document describes only static analysis techniques. 
     \\\\If your suggested confidence level is $>$ 92, the $<response>$ is *YES*. \\If your suggesting confidence is $<$ 85, the $<response>$ is *NO*. \\If your suggested confidence level is $>$ 85 and $<$ 92, the $<response>$ is *DOUBT*. \\You always start your answer by informing the $<response>$, your confidence level in the range of 0-100, and a brief explanation about your decision.   \\
     \hline
    \end{tabularx}
    \caption{Prompt V0.0}
    \label{tab:PromptV0.0}
\end{table}

\begin{table}
    \centering
    \begin{tabularx}{\textwidth}{|X|}
    \hline
     **Context**: You are an expert in context-aware software testing. You must choose software testing documents to support testing context-aware avionics software systems for manned civil aircraft in the industry. You consistently and professionally follow instructions and criteria to support your choice and provide an answer.\\\\

**Instructions**:
1. Clear all of your previous document evaluations.\\
2. Evaluate the documents base on the following 13 rules:\\
  - Rule 1: The document concerns a manned or piloted aircraft.\\
  - Rule 2: The document concerns an aircraft operating within civil aviation.\\
  - Rule 3: The document indicates the aircraft's software.\\
  - Rule 4: The document describes the design, execution, or reporting of the testing of avionics software systems.\\
  - Rule 5: The document describes techniques, technologies, processes, or standards for avionics software testing.\\
  - Rule 6: The document describes the planning, design, execution, or reporting of testing avionics software systems.\\
  - Rule 7: The document describes an application in the industry.\\
  - Rule 8: The document is not an operating or installation manual.\\
  - Rule 9: The document does not describe instruments, equipment, or toolkits to support software testing in general.\\
  - Rule 10: The document does not describe military applications.\\
  - Rule 11: The document does not describe space aircraft or airspace applications.\\
  - Rule 12: The document does not describe formal verification and validation methods.\\
  - Rule 13: The document does not describe static analysis or verification techniques.\\
3. Provide your answer Observing a **Response Criteria** and using an **Output Template**.\\

**Response Criteria**:
Set the `$<choice>$` to "*YES*" if the software testing document satisfies all 13 rules.\\
Set the `$<choice>$` to "*NO*" if the software testing document does not satisfy any of the 13 rules.\\
Set the `$<choice>$` to "*DOUBT*" if you cannot decide based on the rules\\
Justify your decision by filling in an `$<explanation>$` with two short phrases extracted from the software testing document.\\
Set the `$<confidence level>$` with a 0 - 100\% value to indicate your decision confidence.\\
\\
**Output Template**:
`$<choice>$`; "Confidence = "; `$<confidence level>$`; `$<explanation>$`\\
\\
**Examples of Output**:\\
- *YES*; Confidence = 94\%; The document explains how to test context-awareness software testing.\\
- *DOUBT*; Confidence = 91\%; The document regards model-based testing to support the generation of context-awareness test cases.\\
- *NO*; Confidence = 82\%; The document explains how to use formal methods to test software systems.
 \\
     \hline
    \end{tabularx}
    \caption{Prompt V4.1}
    \label{tab:Prompt V4.1}
\end{table}

\section{Empirical observation of the prompt and questions and their ability to screen sources}\label{sec:results}

This section presents the iterative process we followed to evaluate the tools behavior and drive the prompt engineering process. To articulate the purpose of this process we used the Goal-Question-Metric approach, as reported below. 

\begin{tcolorbox}[colframe=blue!50!black, colback=blue!10, coltitle=black, sharp corners=southwest, rounded corners]
\textit{Analyse} an LLM-based filtering component,  and its prompts. \textit{For the Purpose of} characterizing its behavior. \textit{With respect to} its capacity to judge the discard of an information source in a way that is comparable to a human researcher (using the Positive Percent Agreement metric)
\textit{From the viewpoint of} software engineering researchers
\textit{In the context of} supporting the researchers in eliminating irrelevant sources of information for a multivocal literature review regarding the testing of Context-Aware Software Systems of manned aircraft.
\end{tcolorbox}

As previously stated, the goal and use case for the tool is to minimize research bias, or at a minimum, ensure that any introduced bias is comparable to that of a cohesive human research team. As such, this section describes the closed-loop iterative process that was followed during the Prompt-Engineering process. For each change to the \textit{prompt} and/or \textit{question}, we would execute the tool against a suitable sample of PDFs and evaluate an objective metric to understand if the change brought us closer to the goal.

Section \ref{SEC:MonitoringAndMeasuring} presents the rationale for adopting the Positive Percent Agreement metric as our objective metric of choice, we evaluated other, maybe more frequent used, agreement metrics before deciding that the Positive Percent Agreement Metric was the one that most closely aligned with the goals. Following this, Section \ref{sec:validation_in_use} presents quantitative evidence demonstrating the ability of various prompt versions to classify sources without introducing extraneous bias.

\subsection{Evaluating metrics for observing the behavior of the LLM-based tool}\label{SEC:MonitoringAndMeasuring}

To determine whether the behavior of the LLM-based tool would contribute to our goal, we explored inter-rater agreement statistical methods and applied them to assess the progress and capacity of the LLM-based tool in selecting relevant sources.

While this is not the first work to employ LLM-based tools for secondary studies (see Section \ref{sec:RelatedResearch}), a salient feature embedded in our engineering process is the deliberate avoidance of treating researcher judgment as absolute ground truth. We acknowledge that even within a cohesive team like ours, the application of rules such as those stated in the inclusion and exclusion criteria can lead to variations in judgment among team members. 

In short, we required a measure of agreement that: 1) would not rely on a predefined ground truth; 2) would accommodate the inherent variability in judgment; and 3) would specifically focus on evaluating the rejected papers (i.e., \textbf{No} votes).

Consequently, we explored several agreement statistics methods to guide the development of our tool:

\begin{itemize}
    \item \textbf{Direct Perceptual Agreement}: This can be calculated as the overall agreement between two raters. However, this straightforward measure assumes that one rater possesses the ground truth and does not account for randomness or variation in the decision. Therefore, we did not consider it suitable for our engineering process.
    \item \textbf{Cohen's Kappa} \cite{Cohen_Kappa}: This statistic quantifies the agreement between two raters, considering all possible judgment categories and adjusting for the possibility of random agreement. Empirically, an agreement level above 30\% generally indicates genuine agreement beyond chance. While we initially considered Cohen's Kappa, we discarded its use because it directly compares two raters and considers all voting categories (in our case: Yes, No, Doubt). Our tool's primary goal, however, was to discard sources that would not be relevant to our research (i.e., \textbf{No} votes only).
    \item \textbf{Fleiss's Kappa} \cite{fleiss_Kappa}, While we used Fleiss' Kappa to demonstrate that our team voting was consistent and exceeded the threshold of randomness, this metric is not suitable for evaluating the output from the LLM-Based tool, as it is designed for multi-rater agreement within a group, not for assessing an individual system's performance.
    \item \textbf{Positive Percent Agreement} (PPA), \cite{Fletcher2005_PPA}, is a statistical measure that quantifies the proportion of positive cases correctly identified by a particular assessment method. The PPA specifically focuses on the agreement within the positive category. While its primary application is often seen in evaluating the efficacy of diagnostic tests in correctly identifying the presence of a condition, its underlying focus on a specific outcome category also applies to our context of evaluating agreement on 'No' votes.
    While the PPA directly measures agreement on the positive category, it implicitly acknowledges the potential for variation from other output categories. Therefore, by focusing on the PPA for the \textbf{No} category, we gain a targeted metric that reflects the consistency with which our human reviewers agreed on the irrelevance of sources. A higher PPA in this context signifies a greater degree of confidence that the LLM tool is effectively replicating the reviewers' ability to identify and discard irrelevant sources correctly.
\end{itemize}

\subsection{Validation: Quality in use}\label{sec:validation_in_use}

This section presents the evolution of the PPA metric as we iterated through the development stages of the tool outlined in the previous sections. As we mentioned before, a key design consideration was the constrain of utilising readily available hardware within our research team (and not relying on API-based solutions). To illustrate this, the following hardware platforms were used for evaluating the different versions of the \textit{prompt} and \textit{question}:

\begin{itemize}
    \item \textbf{Platform 1 / Laptop with low-end GPU:} Intel Core i7-EVO, 32 GB RAM, 1 TB SSD. Nvidia T500 4 GB GPU. OS: Windows 11.
    \item \textbf{Platform 2 / GPU Desktop:} Intel Core i7, 32 GB RAM, 1 TB SSD. Nvidia  RTX 4060 8 GB GPU. OS: Windows 11.
    \item \textbf{Platform 3 / Workstation:} 2× Intel Xeon E5-2650 v3 2.30GHz, 192 GB RAM, 3 TB SSD. Nvidia  RTX 4090 24 GB GPU. OS: Linux.
\end{itemize}

It is important to clarify that the presentation of these platforms is not intended as a performance benchmark, but rather to demonstrate the feasibility of deploying LLM tools on relatively low-end hardware platforms. Table \ref{tab:SampleAndPurpose} coveys how, as our confidence in the tool grew, we also created more representative statistical samples of the entire dataset. The decision to run larger datasets on more powerful hardware aimed to reduce the turnaround time between each execution. Table \ref{tab:PromptEvo} presents in which platform each dataset was executed. The data in the Processing time column indicates the execution time it took the team to obtain feedback from the LLM on each platform.

\begin{table}[h]
    \centering
    \begin{tabularx}{\textwidth}{XXXXX}
        \textbf{Sample name} & Development & Evaluation & Validation & Full \\
        \hline
        Purpose & Programming & Programming and Prompt Engineering & Validation in Use & Continuing research agenda\\
        Sampling Method & Convenience Sampling& Random / not Representative & Statistical Random / Representative & Full dataset \\
        Size & Small 1 to 10 PDFs & Medium  58 PDFs&  Large 368 PDFs & Full dataset 8482 PDFs \\ 
    \end{tabularx}
    \caption{Definition, sampling method, and purpose of each sample dataset}
    \label{tab:SampleAndPurpose}
\end{table}

\begin{table}[h]
    \centering
    \begin{tabularx}{\textwidth}{p{1.4cm}Xp{1.3cm}XX}
    \hline
    \textbf{User Question} & \textbf{Prompt Version / Releases}\footnote{
   This column identifies the various executions the LLM-based tool underwent during development and prompt engineering. While the naming strategy was generally a Major.Minor convention, semantic information was occasionally added to version names during exploratory phases, as exemplified in Fig. \ref{Fig_PPA_Evolution}. For a detailed account of all prompt versions, readers are referred to the replication package.} & \textbf{Platform}  & \textbf{Sample} & \textbf{Processing Time (min)} \\ \hline
    UQ0 & 0.0-0.4 & 1 & Development & $<20$ \\ 
    UQ1 & 1.0-1.8 & 1, 2 & Development, \makecell{Evaluation} & In Platform 2: \makecell{$<5$ (Development),} \makecell{$<18$ (Evaluation)} \\ 
    UQ2 & 2.0-2.3 & 2, 3 & Evaluation, \makecell{Validation} & In Platform 3: \makecell{$<6$ (Evaluation),\\$<23$ (Validation)} \\ 
    UQ3 & 3.0-3.1 & 2, 3 & Evaluation, \makecell{Validation} & In Platform 3: \makecell{$<6$ (Evaluation),\\ $<23$ (Validation)} \\ 
    UQ4 & 4.0-4.1 & 3 & Validation, Full & $<23$ (Validation), $<2160$ (36 hours) (Full) \\ \hline
    \end{tabularx}
    \caption{Details of User Questions and Processing Times} 
    \label{tab:PromptEvo}
\end{table}

\subsubsection{Benchmark agreement of Inclusion and Exclusion criteria for the avionics section}

As mentioned in section \ref{sec:validation_in_use}, we iteratively sampled from the full dataset to ensure each sample was more representative (see Table \ref{tab:SampleAndPurpose}).

To apply the PPA statistics within our engineering process, we drew a random sample of \textit{58} sources, which is the \textit{Evaluation} sample in Table \ref{tab:SampleAndPurpose}. The sources in the \textit{Evaluation} sample were then independently assessed by three researchers on our team, who assigned one of three votes: Yes (indicating relevance), Doubt (indicating uncertainty), or No (indicating irrelevance). Subsequently, we employed a predefined aggregation protocol \footnote{For instance, a source was classified as "Included" if all three researchers voted "Yes," or if at least two voted "Yes" and the third voted "Doubt." A unanimous "Doubt" resulted in a "Doubt" classification. If exactly two researchers voted "Doubt" and the third voted either "Yes" or "No," the outcome was also "Doubt." All other voting combinations resulted in a "No" classification.}  to determine the final inclusion status of each source. This process yielded a dataset of \textit{58} sources, each classified by three expert researchers, representing a plausible set that could have progressed to the subsequent stage of our MVLR protocol. As an internal measure of inter-rater reliability for the initial voting, we calculated Fleiss' Kappa, which yielded a value of \textbf{0.49}, indicating moderate agreement among the researchers. And both confirm how, even within a cohesive team, including and excluding criteria are subject to varying interpretations. 

\begin{figure}[ht]
\centering
\includegraphics[width=\textwidth]{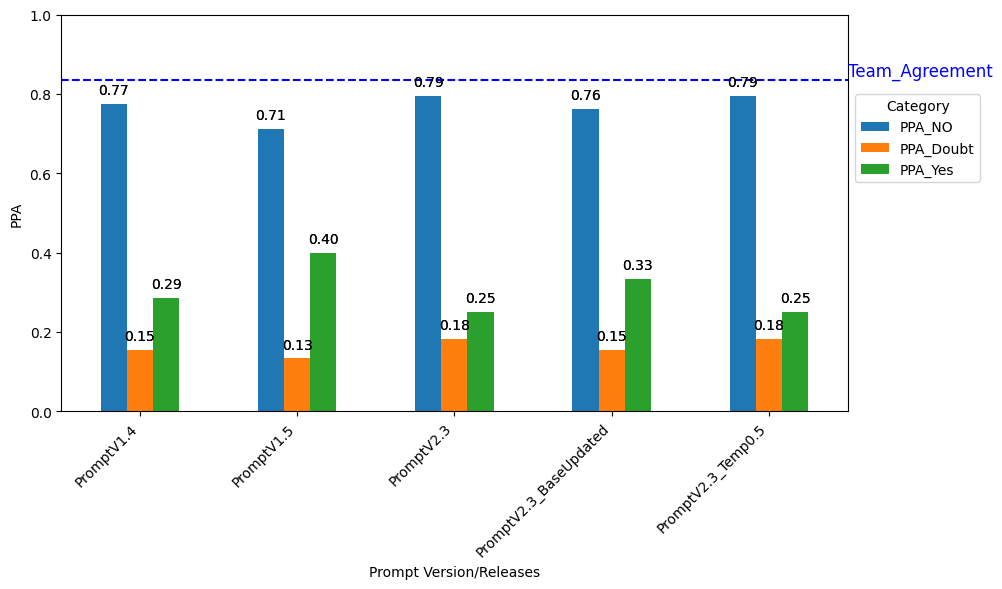}
\caption{PPA progression across versions with the 58 PDFs in the \textit{Evaluation} sample}
\label{Fig_PPA_Evolution}
\end{figure}

This dataset of \textit{58} sources was internally designated as \textbf{TeamAgreement} and served as the input for calculating the PPA for all iterations of the LLM-based tool. Specifically, following each iterative development or modification of the tool (as detailed above), we computed the PPA to evaluate its performance. An increasing PPA value indicated a progressive alignment of the tool's output with the consensus reached by our research team. While the guiding metric is the PPA(No) metric, we calculated and presented the PPA for the other two voting options. Figure  \ref{Fig_PPA_Evolution} presents the results of the selected version through the iterative process. 

A local maximum was observed with version PromptV2.3, yielding a \(PPA(No)=0.79\) . This result led to subsequent analysis and explorations of the effects of code changes (PromptV2.3\_BaseUpdated) and temperature adjustments (PromptV2.3\_Temp0.5). The outcomes, shown in Fig. \ref{Fig_PPA_Evolution}, indicate some performance variation across these attempts, yet none achieved the target metric.

Furthermore, we reviewed all the collected results in search of voting patterns resulting from the different versions of the LLM. With this analysis, we were able to identify a consistent voting pattern by the LLM across different iterations for a subset of sources. For these sources, the LLM had consistently maintained the same Vote through the other iterations.  This analysis also required us to review the sources while interacting with the LLM through the different prompt versions. Overall, we started to realized that the interaction with the LLM was shaping our collective understanding of the Inclusion and Exclusion Criteria (IC/EC). 

\begin{figure}[ht]
\centering
\includegraphics[width=\textwidth]{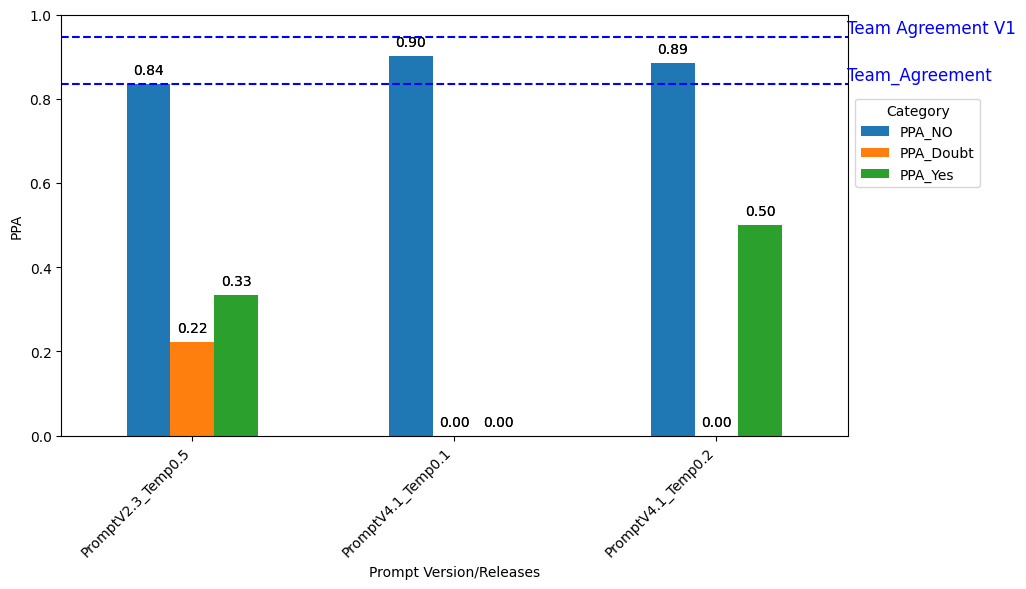}
\caption{PPA progression across versions after reevaluating the 58 PDFs in the \textit{Evaluation} sample}
\label{Fig_PPA_Evolution_after_feedback}
\end{figure}

Consequently, we selected a sample of \textbf{nine} papers from the \textbf{Evaluation} dataset. This sample intentionally included sources where the LLM had exhibited consistent voting alongside a random selection of other sources. The same three researchers were then asked to re-evaluate their initial votes for these \textbf{nine} papers, without being informed of the LLM's voting consistency on any PDFs with a voting that differed from the \textbf{TeamAgreement}. This review process resulted in a revised consensus for the original  \textbf{58 sources in the \textit{Evaluation} sample}, which we named \textbf{TeamAgreement\_V1}. This refined dataset served as the benchmark against which the final versions of the LLM were evaluated. In Fig. \ref{Fig_PPA_Evolution_after_feedback}, the final version of the tool is shown in comparison with \textbf{TeamAgreement\_V1}. The reader will note the difference in performance for version \textit{PromptV2.3\_BaseUpdated} between Fig \ref{Fig_PPA_Evolution} and Fig \ref{Fig_PPA_Evolution_after_feedback}, conveying how our collective decisions regarding the 58 PDFs in the \textit{Evaluation} changed without interaction with the LLM (we expand on the implications of this in section \ref{sec:Discussion}).

The results for \textbf{PromptV4.1\_Temp0.1} warrant specific discussion. While this version attained the highest \textit{PPA(No)} score, it failed to register any agreement for \textit{PPA(Yes)} or \textit{PPA(Doubt)}. Upon inspection, we found that all sources were classified as either 'Yes' or 'No', resulting in a single strong disagreement between the execution and TeamAgreementV1. Although we hypothesize this deterministic behavior is due to the low temperature setting, it serves as an important indicator of the potential risks in automated source selection. In this case, one source out of the 58 in the Evaluation sample would have been discarded (potentially erroneously) before subsequent MVLR stages. This scenario, explored further in Section \ref{sec:Discussion}, represents an example of the fundamental trade-off that motivated this research: weighing the value of saved time and effort against the research risk posed by potential false negatives.

\subsubsection{Evaluation of the LLM version against the Large representative sample}
The process described in the previous section provided sufficient confidence that the LLM's votes were consistent with those of the researchers. To evaluate this consistency more rigorously, we drew a statistically representative sample from the total population of identified sources. For a population size of 8482 potential sources, a sample size of 368 was determined to be necessary to achieve a 95\% confidence level with a 5\% margin of error, assuming a population proportion of 50\% \footnote{\href{https://www.calculator.net/sample-size-calculator.html}{Sample Size Calculator at Calculator.net}}. This sample was then processed by the LLM-based tool, and each of the three researchers was assigned 73 or 74 sources for individual assessment. The Positive Percent Agreement (PPA) calculated between the LLM's votes and the aggregated votes of the researchers for this sample was \textbf{96\%}, indicating a very high degree of consistency between the manual review process and the LLM-based approach.

The distribution of votes for this representative sample is detailed in Table \ref{tab:Sample368_votes}

\begin{table}[h]
    \centering
    \begin{tabular}{lcccc}
        \toprule
        Category & Doubt & Yes & N/A & No \\
        \midrule
        LLM Vote & 0 & 11 & 95 & 262 \\
        \bottomrule
    \end{tabular}
    \caption{LLM votes and human votes for the representative sample}
    \label{tab:Sample368_votes}
\end{table}

\subsection{Final run for research}\label{Sed:fullRun}
\begin{figure}[ht]
\centering
\includegraphics[scale=0.6]{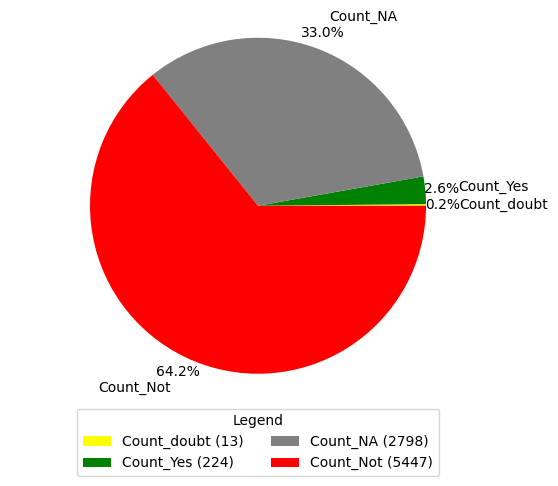}
\caption{Distribution of the final run}
\label{Fig_FinalRunPieChart}
\end{figure}

The final step was submitting all identified sources to the LLM-based tool. This run will allow us to continue our research into CASS Testing. The results are presented in Table \ref{tab:FinalDistribution} and Fig. \ref{Fig_FinalRunPieChart}.

\begin{table}[h]
    \centering
    \begin{tabular}{lcccc}
        \toprule
        Category & Doubt & Yes & N/A & Not \\
        \midrule
        LLM Vote & 13 & 224 & 2798 & 5447 \\
        \bottomrule
    \end{tabular}
    \caption{Distribution of final votes by the team and LLM-Tool votes}
    \label{tab:FinalDistribution}
\end{table}

\section{Discussion}\label{sec:Discussion}
Our iterative engineering process for the LLM-based tool yielded several noteworthy insights, which are discussed in detail below.

\textbf{Prompt Engineering}. Engineering involves applying knowledge to solve problems within a specific domain. In the context of Prompt Engineering, however, the knowledge base remains relatively limited. As detailed in Section \ref{Sec_PETheory}, a literature review was conducted to identify best practices for defining and improving our prompts. Nevertheless, when evaluating the evolution of these prompts and their impact on our chosen metric (PPA) using a controlled set of \textbf{58} source samples, the process appeared significantly driven by trial and error. This observation suggests that current \textit{Prompt Engineering} practices lack a robust foundation of established knowledge. Despite this, our analysis did reveal several potentially valuable insights for the further development of this practice. These insights are derived from observing the versions of the LLM-based tool that demonstrated the most substantial effect on our guiding metric, as well as the tweaks that did not affect the model's performance against our goal.

\textbf{The effects of ambiguous positive and negative statements in the Prompt confused the LLM model.} A significant improvement in performance during our development process occurred when the \textit{exclusion criteria} were rephrased as direct statements. Table \ref{tab:PromptComparison} illustrates this difference. Initially, the prompt was structured similarly to typical inclusion/exclusion criteria. However, a shift towards direct and assertive statements yielded substantially better model performance. As shown in the \textit{after} column of Table \ref{tab:PromptComparison}, the Exclusion Criteria concerning operations manuals was rewritten, aligning with the instruction ``Evaluate the documents
base on the following 13 rules''. We hypothesize that the model is confused by the instruction with a negative outlook (i.e., the word \textit{reject}), yet it receives a set of affirmations (i.e., \textit{the document is}).

\begin{table}[h]
    \centering
    \begin{tabularx}{\textwidth}{X X}
        \toprule
        \textbf{Before} & \textbf{After} \\
        \midrule
        ...You always \textit{reject} a document when any of these five rules are satisfied:  \newline
        1 - The document is an Operating or installation manual;... &  
        You always select a document when all of these eleven rules are satisfied:  \newline
        ...  \newline
        7 – The document is not an Operating or installation manual; \\
        \bottomrule
    \end{tabularx}
    \caption{Prompt comparison between positive and negative statements}
    \label{tab:PromptComparison}
\end{table}

We note that determining the cause of this behavior is not within the scope of our research. Furthermore, our concept of performance improvement, specifically concerning the prompt and output, is closely tied to the LLM tool's ability to reject irrelevant papers. Nonetheless, we include this observation to contribute to the growing, yet still preliminary, body of evidence surrounding prompt engineering approaches. We argue that this reinforces our earlier point regarding the current state of prompt engineering, which has not yet developed the characteristics of a mature engineering discipline.

\textbf{Template Prompt for other teams using our tool}
As mentioned, the tool development closely followed the restrictions in the research protocol. However, we have discussed at length what would be needed to change to use the tool in another research domain. We leave here a proposed prompt template for others to use our tool in their research. This prompt is based on our lessons learned and our current prompt, and adopting the presentation from \cite{xu2025_expertprompting}:

\begin{tcolorbox}[colframe=blue!50!black, colback=blue!10, coltitle=black, sharp corners=southwest, rounded corners]
[Expert] You are a \{Role\} specialized in selecting documents talking about \{Subject\}.

[Instruction] You always select a document when all of these \{Number of Inclusion Criteria\} rules are satisfied: \\
\{Numbered list of Inclusion Criteria\} 1 – \{Inclusion Criterion\}; 2 - \{Inclusion Criterion\}; \{n\} - \{Inclusion Criterion\}. 

[Instruction] You always reject a document when any of these \{Number of Inclusion Criteria\} rules are satisfied:
\{Exclusion Criterion\} 1 – \{Exclusion Criterion\}; 2 - \{Exclusion Criterion\}; {n} - \{Exclusion Criterion\}.

[One Shot Answer] If your suggested confidence level is $>$ 92, $<$response$>$ is set to '*YES*'. If your suggested confidence is $<$ 85, $<$response$>$ is set to '*NO*'. If your suggested confidence level is  $>$ 85 and $<$ 92, $<$response$>$ is set to '*DOUBT*'.

[Instruction] You always start your answer by informing the $<$response$>$, your confidence level in the range of 0-100\%, and a brief explanation about your decision.
\end{tcolorbox}

\textbf{Regarding Model Selection, Prompts, and Model Benchmarking.} The primary focus of our research lies in CASS testing; consequently, benchmarking the performance of different LLM models falls outside our immediate scope. Our engagement with LLMs is that of users leveraging an available technology. As such, our initial model selection was primarily dictated by the computational resources available to run a model locally. Therefore, the key criterion for choosing \textit{Dolphin-llama3}\footnote{https://ollama.com/library/dolphin-llama3} was its feasibility for local deployment on our workstations.

Throughout our experimentation with this model and the iterative process of \textit{Prompt Engineering}, we hypothesize that there was a strong interdependency between the prompt design and the specific LLM employed. Our repeated observations suggested that simply transitioning to a newer model within the same family would not necessarily yield improved performance without corresponding adjustments to the prompt. To illustrate this, we conducted a small exploratory experimentation. For this, we used the  \textbf{58} source samples and made no changes to the source code, maintaining the same values for parameters such as temperature (set at 0.1), as well as the question and prompts. We chose three models that a search showed were variations of the one we used throughout the development. These models are: \textbf{llama3} in its 8B parameter version, \textbf{llama3.1} in its 8B parameter version, and the latest updated version of \textbf{dolphin-llama3}, also in its 8B parameter version (which is safe to assume is the most similar model to the one used throughout development).

\begin{figure}[ht]
\centering
\includegraphics[width=\textwidth]{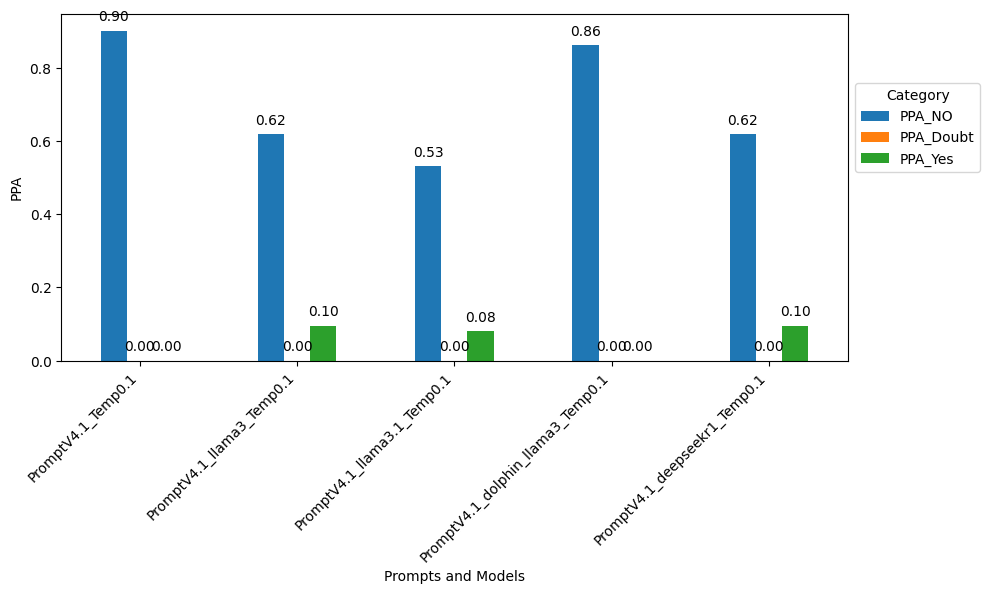}
\caption{PPA for the execution of the tool with different underlying models without changes to Prompt or Question}
\label{Fig_PPA_OtherModels}
\end{figure}

Our results demonstrate that the performance of the tool varies significantly depending on the underlying model used for the judgments (see Figure \ref{Fig_PPA_OtherModels}). For researchers seeking to utilize our tool, these results reinforce the argument that the question and prompt must be carefully evaluated and tailored to the research domain to ensure that the tool's results do not introduce unnecessary bias into the selection process.

Similar to the previous point, our observations regarding prompt engineering and model behavior are shared with the community to promote the advancement of LLMs as supportive tools in specific application domains. These observations should not be interpreted as a comprehensive analysis or benchmarking of LLM capabilities. Further research focusing on the nuances of prompt engineering methodologies and comparative model performance would be necessary to draw broader conclusions.

\textbf{Processing time and hardware constrains}. The overall processing time for the full sample is related to the underlying hardware. However, our intention in experimenting with different underlying platforms stems from pragmatism rather than optimization. As mentioned previously, our choice of foundational LLM model was driven and restricted by the capacity of our workstations to execute it locally. Since each source (i.e., PDF) is processed individually, we were able to run the sample with the hardware we had access to.

We also noted that the execution time of some steps remained fairly constant regardless of the underlying execution platforms. For example, with minor variations to account for network conditions, downloading each PDF takes roughly the same amount of time on any of the platforms.

Improvements in hardware are most noticeable in the inference request to the LLM. 

\textbf{Optimization, state-of-the-practice RAG, and fit for purpose of the LLM-tool} 
As mentioned in \ref{sec:RAG_Encoding}, we encoded the PDF content into plain text before sending it to the LLM for processing. We acknowledge that Retrieval Augmented Generation (RAG) best practices evolved rapidly during our development cycle. For instance, current best practices often recommend encoding sources into vector databases optimized for large language model (LLM) processing.

While this alternative approach could potentially improve execution times, our primary goal was to demonstrate that the tool could free up researchers' time without injecting bias. Given the available hardware, the execution times were acceptable, allowing us to prioritize validating the tool and ensuring the quality of its responses for our research purposes.

\subsection{Limitations and threats to validity}\label{sec:Limitations}
In this section, we discuss several limitations and threats to the validity of using the tool to support research.

\textbf{Effects of feedback and auto-influence.} A potential threat to the construct validity of our study lies in the evolving understanding of the research domain, the quality and content of the available sources, and the application of \textit{inclusion and exclusion criteria} throughout the iterative prompt engineering process. As we refined the prompts and reviewed the model's output on different samples, we engaged in discussions. This led to a shift in our comprehension of the sources, and the interpretation of the inclusion and exclusion criteria inevitably changed. This creates a reinforcing loop: our developing knowledge informs subsequent prompt iterations, and the results from those iterations, in turn, further shape our understanding. This reinforced loop, where the researchers are active participants within the development and evaluation cycle, introduces a degree of subjectivity that is difficult to mitigate entirely. Our evolving understanding directly influenced the model's development, and conversely, the model's performance provided feedback that refined our domain knowledge.

\begin{figure}
    \centering
    \begin{minipage}{0.4\textwidth}
        \centering
        \includegraphics[width=\linewidth]{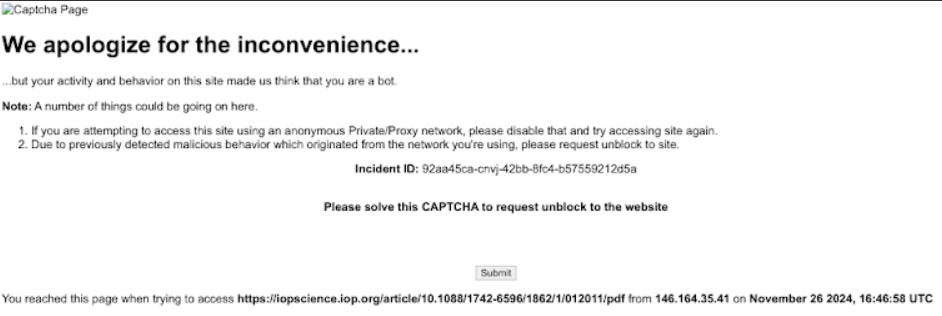}
        (a) Screenshot of the PDF as downloaded by the tool
    \end{minipage}
    \hfill
    \begin{minipage}{0.4\textwidth}
        \centering
        \includegraphics[width=\linewidth]{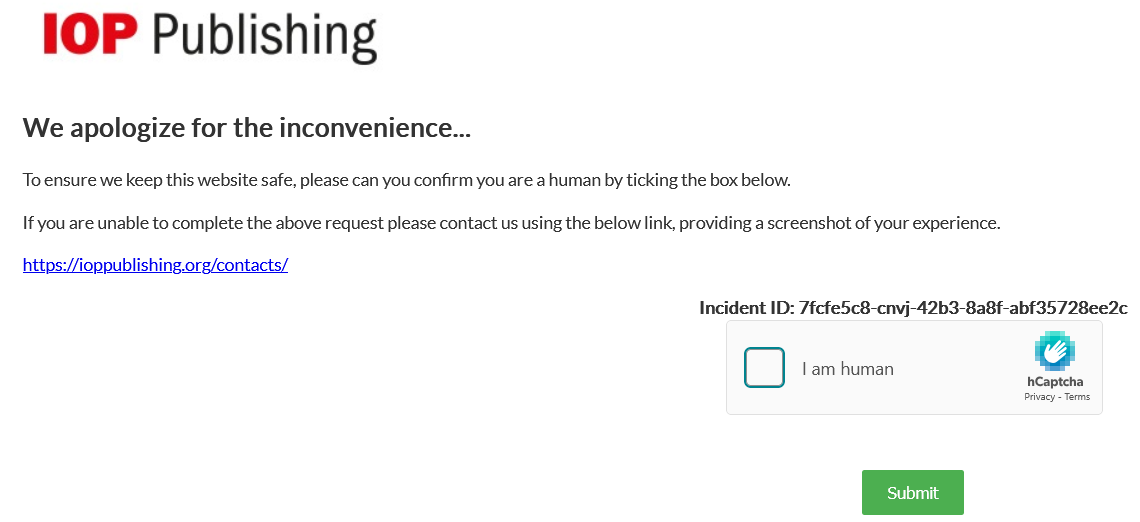}
        (b) Screenshot of the Captcha page when a human access the URL
    \end{minipage}
    \hfill
    \begin{minipage}{0.4\textwidth}
        \centering
        \includegraphics[width=\linewidth]{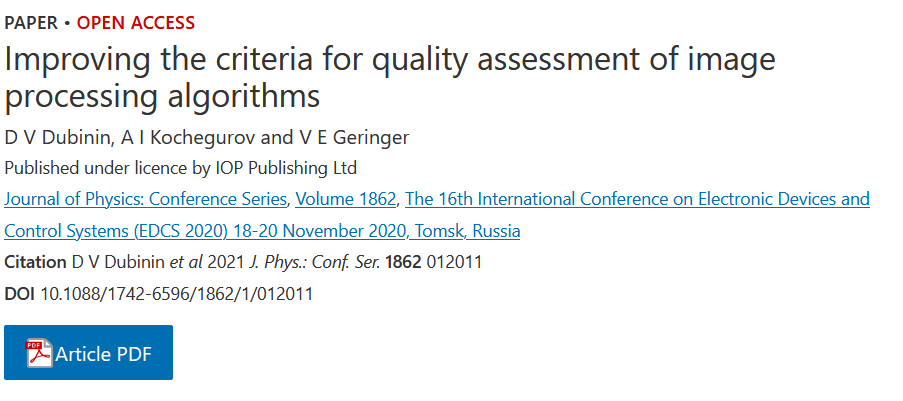}
        (c) Screenshot of source web page \cite{CaptchaPaper_Dubinin_2021} when a human completes the captcha
    \end{minipage}
    \label{Fig:CaptchaEffect}
    \caption{Images (a), (b), and (c) show the effects of the Captcha on the Automated script (a), and the different content that a human can process (b) and (c).}
\end{figure}

\textbf{Sampling and effects of Distributed Denial of Service (DDoS) Gateways} . A potential threat to the external validity of our findings arises from limitations in data acquisition due to website access restrictions, such as DDoS protection gateways. We observed instances where our automated scraping process was blocked, resulting in the retrieval of PDF documents containing error messages (see "Count\_NA" in Figure \ref{Fig:CaptchaEffect}). When processed by our script and the LLM, these PDFs often led to a 'No' judgment. As the content downloaded in the PDF is not relevant to the research. However, this classification does not necessarily reflect the actual relevance of the source, as the LLM tool could not access and process the \textit{content} that was indexed by Google and for which the link was initially generated.

Human reviewers, in contrast, could identify these 'Not Available' cases. Our analysis of the 368-source sample revealed that all instances marked as 'Not Available' by the researchers corresponded to inaccessible content for the LLM (i.e., true unavailability). However, the converse was not always true; the LLM could not definitively determine the underlying reason for a 'Not Available' response.

In our 368-source sample, this issue accounted for approximately 10\% of the reviewed links. Assuming a similar proportion within the full population of 8482 potential sources, many false negatives (irrelevant classifications due to access issues) could be present in the LLM's output. Identifying these true 'Not Available' cases would necessitate a manual inspection of the 'No' classifications, partially undermining the intended efficiency gains of the automated approach.

Consequently, the acceptance of these potential false negatives represents a trade-off made to accelerate the overall process of automatically reviewing a large volume of material. While this limitation may introduce some degree of error, it is a necessary compromise for substantially reducing the manual effort required for the secondary study.

\textbf{Reproducibility of the research and effects of the guiding MVLR protocol}. Several factors contribute to the rigor and potential for replication of our work. Firstly, our LLM-based tool has been specifically developed and tuned for our particular research domain within CASS testing. Secondly, we have made the data and source code used in our analysis publicly available to enhance transparency and facilitate scrutiny.
The central principle underpinning the rigor and relevance of our tool for its intended purpose – supporting our main research aims in CASS testing – is the strict adherence to our established research protocol in every step automated within the LLM-based tool.
We argue that researchers seeking to replicate our results can readily utilize the provided source code for the tool. However, as discussed in the preceding section, the prompt requires careful tuning to align with the specific domain's inclusion and exclusion criteria, as well as the selected LLM model. Furthermore, the successful application and validation of such a tool necessitate a foundation of sound experimental practices, exemplified by the research protocol we employed.

\section{Conclusion}\label{sec:Conclusion}
This paper presents the development and validation of an LLM-based tool designed to support researchers in executing an MVLR on the topic of Context-Aware Software System testing for the avionics industry.

Throughout our engineering process, we prioritized that the tool should not pose a threat to the execution of the MVLR. At the very least, whatever bias the tool was intended to introduce should be comparable to the bias introduced by human researchers. As the task entrusted to this tool was to discard sources that would not be relevant to the topic of interest of the MVLR, we turned to Agreement statistical methods to guide the engineering of the tool. We incorporated the notion that even within cohesive teams, agreement cannot be perfect, and therefore, a degree of variation must be accounted for.  The tool shows high levels of agreement with researchers of \textbf{\GetValue{FinalPPAITeration2}}. 

To further assure this, the engineering process iteratively and systematically sampled the potential sources  and incrementally evaluated the tool with the different samples. This increased our confidence that the tool would perform as intended when used to advance our scientific interest in context-aware testing.  As a result, we presented a tool capable of processing at least 8,482 sources, and we are confident in its results, which will inform our continued research on CASS (section \ref{Sed:fullRun}). Thereby releasing precious human time without introducing bias into the selection process. 

In addition, this paper highlights several key observations that are important when considering the use of the presented tool, or any LLM-based tool, for scientific research. Specifically, during the prompt engineering phase, we noticed that refining the prompt to enhance the tool's reliability created a reinforcing loop. This process, in turn, reshaped our understanding of our inclusion and exclusion criteria. This created another  feedback loop, as the development of the tool helped us focus on our target and avoid false positives. 

Finally, we have emphasized that the tool's development relied on a rigorous research protocol. This protocol, which has supported previous MVLR, guided our decisions and established the operational boundaries for the LLM-based tool. Researchers looking to reproduce the results of this paper should also establish a research protocol and ensure a clear understanding of their inclusion and exclusion criteria. As we firmly believe that it is human researchers who must guide the direction of the LLM-based assistant, the rigor, relevance, and importance of the results are strongly grounded in sound research practices.

\section{Statements and Declarations}

\begin{itemize}
\item Funding: Prof. Travassos is  a Brazilian Research Council (CNPq) Researcher (grant 305701/2022-3) and a State Scientist FAPERJ – Carlos Chagas Filho Foundation for Research Support of the State of Rio de Janeiro (grant E-26/204.310/2024)   
\item Conflict of interest: The authors declare that they have no conflict of interest.
\item Data availability: All data used to write this paper are open-source and available at: \href{https://git-lab.cos.ufrj.br/contextaware/llm/}{https://git-lab.cos.ufrj.br/contextaware/llm/}
\item Code availability: All code used to write this paper is open source and available at: \href{https://git-lab.cos.ufrj.br/contextaware/llm/}{https://git-lab.cos.ufrj.br/contextaware/llm/}
\item Author contribution: All authors have contributed equally to the design, methodology, and writing of this manuscript.
\end{itemize}

\bibliography{sn-bibliography}


\begin{thebibliography}{43}
\ifx \bisbn   \undefined \def \bisbn  #1{ISBN #1}\fi
\ifx \binits  \undefined \def \binits#1{#1}\fi
\ifx \bauthor  \undefined \def \bauthor#1{#1}\fi
\ifx \batitle  \undefined \def \batitle#1{#1}\fi
\ifx \bjtitle  \undefined \def \bjtitle#1{#1}\fi
\ifx \bvolume  \undefined \def \bvolume#1{\textbf{#1}}\fi
\ifx \byear  \undefined \def \byear#1{#1}\fi
\ifx \bissue  \undefined \def \bissue#1{#1}\fi
\ifx \bfpage  \undefined \def \bfpage#1{#1}\fi
\ifx \blpage  \undefined \def \blpage #1{#1}\fi
\ifx \burl  \undefined \def \burl#1{\textsf{#1}}\fi
\ifx \doiurl  \undefined \def \doiurl#1{\url{https://doi.org/#1}}\fi
\ifx \betal  \undefined \def \betal{\textit{et al.}}\fi
\ifx \binstitute  \undefined \def \binstitute#1{#1}\fi
\ifx \binstitutionaled  \undefined \def \binstitutionaled#1{#1}\fi
\ifx \bctitle  \undefined \def \bctitle#1{#1}\fi
\ifx \beditor  \undefined \def \beditor#1{#1}\fi
\ifx \bpublisher  \undefined \def \bpublisher#1{#1}\fi
\ifx \bbtitle  \undefined \def \bbtitle#1{#1}\fi
\ifx \bedition  \undefined \def \bedition#1{#1}\fi
\ifx \bseriesno  \undefined \def \bseriesno#1{#1}\fi
\ifx \blocation  \undefined \def \blocation#1{#1}\fi
\ifx \bsertitle  \undefined \def \bsertitle#1{#1}\fi
\ifx \bsnm \undefined \def \bsnm#1{#1}\fi
\ifx \bsuffix \undefined \def \bsuffix#1{#1}\fi
\ifx \bparticle \undefined \def \bparticle#1{#1}\fi
\ifx \barticle \undefined \def \barticle#1{#1}\fi
\bibcommenthead
\ifx \bconfdate \undefined \def \bconfdate #1{#1}\fi
\ifx \botherref \undefined \def \botherref #1{#1}\fi
\ifx \url \undefined \def \url#1{\textsf{#1}}\fi
\ifx \bchapter \undefined \def \bchapter#1{#1}\fi
\ifx \bbook \undefined \def \bbook#1{#1}\fi
\ifx \bcomment \undefined \def \bcomment#1{#1}\fi
\ifx \oauthor \undefined \def \oauthor#1{#1}\fi
\ifx \citeauthoryear \undefined \def \citeauthoryear#1{#1}\fi
\ifx \endbibitem  \undefined \def \endbibitem {}\fi
\ifx \bconflocation  \undefined \def \bconflocation#1{#1}\fi
\ifx \arxivurl  \undefined \def \arxivurl#1{\textsf{#1}}\fi
\csname PreBibitemsHook\endcsname

\bibitem[\protect\citeauthoryear{Garousi et~al.}{2016}]{garousi_MVLR_2016}
\begin{bchapter}
\bauthor{\bsnm{Garousi}, \binits{V.}},
\bauthor{\bsnm{Felderer}, \binits{M.}},
\bauthor{\bsnm{Mäntylä}, \binits{M.V.}}:
\bctitle{The need for multivocal literature reviews in software engineering}.
In: \bbtitle{Proceedings of the 20th {International} {Conference} on {Evaluation} and {Assessment} in {Software} {Engineering} - {EASE} '16},
pp. \bfpage{1}--\blpage{6}.
\bpublisher{ACM Press},
\blocation{New York, New York, USA}
(\byear{2016}).
\doiurl{10.1145/2915970.2916008}
\end{bchapter}
\endbibitem

\bibitem[\protect\citeauthoryear{Matalonga et~al.}{2025}]{DreamTeam_TII_2025}
\begin{barticle}
\bauthor{\bsnm{Matalonga}, \binits{S.}},
\bauthor{\bsnm{Amalfitano}, \binits{D.}},
\bauthor{\bsnm{Solari}, \binits{M.}},
\bauthor{\bsnm{Rossa~Hauck}, \binits{J.C.}},
\bauthor{\bsnm{Travassos}, \binits{G.H.}}:
\batitle{Testing context-aware software systems from the voices of the automotive industry}.
\bjtitle{IEEE Transactions on Industrial Informatics}
\bvolume{21}(\bissue{5}),
\bfpage{1551}--\blpage{3203}
(\byear{2025})
\doiurl{10.1109/TII.2025.3529918}
\end{barticle}
\endbibitem

\bibitem[\protect\citeauthoryear{Matalonga et~al.}{2017}]{Matalonga2017}
\begin{barticle}
\bauthor{\bsnm{Matalonga}, \binits{S.}},
\bauthor{\bsnm{Rodrigues}, \binits{F.}},
\bauthor{\bsnm{Travassos}, \binits{G.H.}}:
\batitle{Characterizing testing methods for context-aware software systems: {Results} from a quasi-systematic literature review}.
\bjtitle{Journal of Systems and Software}
\bvolume{131},
\bfpage{1}--\blpage{21}
(\byear{2017})
\doiurl{10.1016/j.jss.2017.05.048}
\end{barticle}
\endbibitem

\bibitem[\protect\citeauthoryear{Santos et~al.}{2017}]{Santos2017}
\begin{barticle}
\bauthor{\bsnm{Santos}, \binits{I.d.S.}},
\bauthor{\bsnm{Andrade}, \binits{R.M.d.C.}},
\bauthor{\bsnm{Rocha}, \binits{L.S.}},
\bauthor{\bsnm{Matalonga}, \binits{S.}},
\bauthor{\bsnm{Oliveira}, \binits{K.M.}},
\bauthor{\bsnm{Travassos}, \binits{G.H.}}:
\batitle{Test case design for context-aware applications: {Are} we there yet?}
\bjtitle{Information and Software Technology}
\bvolume{88},
\bfpage{1}--\blpage{16}
(\byear{2017})
\doiurl{10.1016/j.infsof.2017.03.008}
\end{barticle}
\endbibitem

\bibitem[\protect\citeauthoryear{Siqueira et~al.}{2021}]{siqueira_testing_2021}
\begin{barticle}
\bauthor{\bsnm{Siqueira}, \binits{B.R.}},
\bauthor{\bsnm{Ferrari}, \binits{F.C.}},
\bauthor{\bsnm{Souza}, \binits{K.E.}},
\bauthor{\bsnm{Camargo}, \binits{V.V.}},
\bauthor{\bsnm{Lemos}, \binits{R.}}:
\batitle{Testing of adaptive and context‐aware systems: approaches and challenges}.
\bjtitle{Software Testing, Verification and Reliability}
(\byear{2021})
\doiurl{10.1002/stvr.1772}
\end{barticle}
\endbibitem

\bibitem[\protect\citeauthoryear{Matalonga et~al.}{2022}]{matalonga_alternatives_2022}
\begin{botherref}
\oauthor{\bsnm{Matalonga}, \binits{S.}},
\oauthor{\bsnm{Amalfitano}, \binits{D.}},
\oauthor{\bsnm{Doreste}, \binits{A.}},
\oauthor{\bsnm{Fasolino}, \binits{A.R.}},
\oauthor{\bsnm{Travassos}, \binits{G.H.}}:
Alternatives for testing of context-aware software systems in non-academic settings: results from a {Rapid} {Review}.
Information and Software Technology,
106937
(2022)
\doiurl{10.1016/j.infsof.2022.106937}
\end{botherref}
\endbibitem

\bibitem[\protect\citeauthoryear{{ISO/IEC/IEEE 29119-1:2022}}{2022}]{ISO_29119}
\begin{barticle}
\bauthor{\bsnm{{ISO/IEC/IEEE 29119-1:2022}}}:
\batitle{Software and {Systems} {Engineering} {Software} testing {Part} 1:{Concepts} and definitions}.
\bjtitle{ISO/IEC/IEEE 29119-1:2013}
(\byear{2022})
\doiurl{10.1109/IEEESTD.2022.9698145}
\end{barticle}
\endbibitem

\bibitem[\protect\citeauthoryear{Thompson et~al.}{2024}]{Thompson2024_ISTQB}
\begin{bbook}
\bauthor{\bsnm{Thompson}, \binits{G.}},
\bauthor{\bsnm{Morgan}, \binits{P.}},
\bauthor{\bsnm{Samaroo}, \binits{A.}},
\bauthor{\bsnm{Kurowski}, \binits{J.}},
\bauthor{\bsnm{Williams}, \binits{P.}},
\bauthor{\bsnm{Salmon}, \binits{M.}}:
\bbtitle{Software Testing},
\bedition{5}th edn.
\bpublisher{BCS, The Chartered Institute for IT},
\blocation{Swindon, England}
(\byear{2024})
\end{bbook}
\endbibitem

\bibitem[\protect\citeauthoryear{Matalonga et~al.}{2023}]{DT_Automotive_Protocol_2023}
\begin{botherref}
\oauthor{\bsnm{Matalonga}, \binits{S.}},
\oauthor{\bsnm{Amalfitano}, \binits{D.}},
\oauthor{\bsnm{Solari}, \binits{M.}},
\oauthor{\bsnm{Rossa~Hauck}, \binits{J.C.}},
\oauthor{\bsnm{Travassos}, \binits{G.H.}}:
Testing {Context}-{Aware} {Software} {Systems} in the {Automotive} {Domain}: {A} {Multi} {Vocal} {Literature} {Review} {Protocol} and {Dataset}.
Zenodo
(2023).
\doiurl{10.5281/ZENODO.8346839} .
\url{https://zenodo.org/doi/10.5281/zenodo.8346839}
Accessed 2023-10-26
\end{botherref}
\endbibitem

\bibitem[\protect\citeauthoryear{Waseem et~al.}{2023}]{waseem2023}
\begin{bchapter}
\bauthor{\bsnm{Waseem}, \binits{M.}},
\bauthor{\bsnm{Ahmady}, \binits{A.}},
\bauthor{\bsnm{Liang}, \binits{P.}},
\bauthor{\bsnm{Fehmi}, \binits{M.}},
\bauthor{\bsnm{Abrahamsson}, \binits{P.}},
\bauthor{\bsnm{Mikkonen}, \binits{T.}}:
\bctitle{Conducting systematic literature reviews with chatgpt}.
In: \bbtitle{Proceedings of the 17th International Symposium on Empirical Software Engineering and Measurement (ESEM '23)}.
\bpublisher{ACM}, \blocation{???}
(\byear{2023}).
\doiurl{10.1145/3611322.3611318}
\end{bchapter}
\endbibitem

\bibitem[\protect\citeauthoryear{Alshami et~al.}{2023}]{alshami2023}
\begin{barticle}
\bauthor{\bsnm{Alshami}, \binits{A.}},
\bauthor{\bsnm{Elsayed}, \binits{M.}},
\bauthor{\bsnm{Ali}, \binits{E.}},
\bauthor{\bsnm{Eltoukhy}, \binits{A.E.}},
\bauthor{\bsnm{Zayed}, \binits{T.}}:
\batitle{Harnessing the power of chatgpt for automating systematic review process: Methodology, case study, limitations, and future directions}.
\bjtitle{Systems}
\bvolume{11}(\bissue{7}),
\bfpage{1}--\blpage{7}
(\byear{2023})
\end{barticle}
\endbibitem

\bibitem[\protect\citeauthoryear{Wang et~al.}{2023}]{wang2023}
\begin{bchapter}
\bauthor{\bsnm{Wang}, \binits{S.}},
\bauthor{\bsnm{Scells}, \binits{H.}},
\bauthor{\bsnm{Zuccon}, \binits{G.}}:
\bctitle{Can chatgpt write a good boolean query for systematic review literature search?}
In: \bbtitle{Proceedings of the 46th International ACM SIGIR Conference on Research and Development in Information Retrieval},
pp. \bfpage{1426}--\blpage{1436}
(\byear{2023}).
\doiurl{10.1145/3539618.3591730}
\end{bchapter}
\endbibitem

\bibitem[\protect\citeauthoryear{Guo et~al.}{2024}]{guo2024}
\begin{barticle}
\bauthor{\bsnm{Guo}, \binits{E.}},
\bauthor{\bsnm{Gupta}, \binits{M.}},
\bauthor{\bsnm{Deng}, \binits{J.}},
\bauthor{\bsnm{Park}, \binits{Y.-J.}},
\bauthor{\bsnm{Paget}, \binits{M.}},
\bauthor{\bsnm{Naugler}, \binits{C.}}:
\batitle{Automated paper screening for clinical reviews using large language models: Data analysis study}.
\bjtitle{J Med Internet Res}
\bvolume{26},
\bfpage{48996}
(\byear{2024})
\doiurl{10.2196/48996}
\end{barticle}
\endbibitem

\bibitem[\protect\citeauthoryear{Khraisha et~al.}{2024}]{khraisha2024}
\begin{barticle}
\bauthor{\bsnm{Khraisha}, \binits{Q.}},
\bauthor{\bsnm{Put}, \binits{S.}},
\bauthor{\bsnm{Kappenberg}, \binits{J.}},
\bauthor{\bsnm{Warraitch}, \binits{A.}},
\bauthor{\bsnm{Hadfield}, \binits{K.}}:
\batitle{Can large language models replace humans in systematic reviews? evaluating gpt-4's efficacy in screening and extracting data from peer-reviewed and grey literature in multiple languages}.
\bjtitle{Research Synthesis Methods}
\bvolume{15}(\bissue{4}),
\bfpage{1}--\blpage{11}
(\byear{2024})
\doiurl{10.1002/jrsm.1715}
\end{barticle}
\endbibitem

\bibitem[\protect\citeauthoryear{Robinson et~al.}{2023}]{robinson2023}
\begin{barticle}
\bauthor{\bsnm{Robinson}, \binits{K.A.}}, \betal:
\batitle{Are chatgpt and large language models “the answer” to bringing us closer to systematic review automation?}
\bjtitle{Systematic Reviews}
\bvolume{12}(\bissue{1}),
\bfpage{72}
(\byear{2023})
\doiurl{10.1186/s13643-023-02243-z}
\end{barticle}
\endbibitem

\bibitem[\protect\citeauthoryear{Wilkins et~al.}{2024}]{wilkins2023}
\begin{barticle}
\bauthor{\bsnm{Wilkins}, \binits{J.}}, \betal:
\batitle{A systematic review of chatgpt and other conversational large language models in healthcare}.
\bjtitle{Journal of Medical Internet Research}
\bvolume{26},
\bfpage{22769}
(\byear{2024})
\doiurl{10.2196/22769}
\end{barticle}
\endbibitem

\bibitem[\protect\citeauthoryear{Gupta et~al.}{2023}]{gupta2023}
\begin{barticle}
\bauthor{\bsnm{Gupta}, \binits{B.}},
\bauthor{\bsnm{Mufti}, \binits{T.}},
\bauthor{\bsnm{Sohail}, \binits{S.S.}},
\bauthor{\bsnm{Madsen}, \binits{D.}}:
\batitle{Chatgpt: A brief narrative review}.
\bjtitle{Cogent Business \& Management}
\bvolume{10}(\bissue{1}),
\bfpage{2275851}
(\byear{2023})
\doiurl{10.1080/23311975.2023.2275851}
\end{barticle}
\endbibitem

\bibitem[\protect\citeauthoryear{Alchokr et~al.}{2023}]{alchokr2022}
\begin{bchapter}
\bauthor{\bsnm{Alchokr}, \binits{R.}},
\bauthor{\bsnm{Borkar}, \binits{M.}},
\bauthor{\bsnm{Thotadarya}, \binits{S.}},
\bauthor{\bsnm{Saake}, \binits{G.}},
\bauthor{\bsnm{Leich}, \binits{T.}}:
\bctitle{Supporting systematic literature reviews using deep-learning-based language models}.
In: \bbtitle{Proceedings of the 1st International Workshop on Natural Language-Based Software Engineering}.
\bsertitle{NLBSE '22},
pp. \bfpage{67}--\blpage{74}.
\bpublisher{Association for Computing Machinery},
\blocation{New York, NY, USA}
(\byear{2023}).
\doiurl{10.1145/3528588.3528658}
\end{bchapter}
\endbibitem

\bibitem[\protect\citeauthoryear{Watanabe et~al.}{2020}]{watanabe2020}
\begin{barticle}
\bauthor{\bsnm{Watanabe}, \binits{W.M.}},
\bauthor{\bsnm{Felizardo}, \binits{K.R.}},
\bauthor{\bsnm{Candido}, \binits{A.}},
\bauthor{\bsnm{Ferreira~{de Souza}}},
\bauthor{\bsnm{Campos~Neto}, \binits{J.E.}},
\bauthor{\bsnm{Vijaykumar}, \binits{N.L.}}:
\batitle{Reducing efforts of software engineering systematic literature reviews updates using text classification}.
\bjtitle{Information and Software Technology}
\bvolume{128},
\bfpage{106395}
(\byear{2020})
\doiurl{10.1016/j.infsof.2020.106395}
\end{barticle}
\endbibitem

\bibitem[\protect\citeauthoryear{Girardi et~al.}{2023}]{girardi2023}
\begin{barticle}
\bauthor{\bsnm{Girardi}, \binits{D.}},
\bauthor{\bsnm{Minku}, \binits{L.}},
\bauthor{\bsnm{Cavalcanti}, \binits{A.}},
\bauthor{\bsnm{Ferrara}, \binits{P.}}:
\batitle{Deep learning for software defect prediction: A systematic literature review}.
\bjtitle{Journal of Systems and Software}
\bvolume{199},
\bfpage{111561}
(\byear{2023})
\doiurl{10.1016/j.jss.2023.111561}
\end{barticle}
\endbibitem

\bibitem[\protect\citeauthoryear{Necula et~al.}{2024}]{necula2024}
\begin{barticle}
\bauthor{\bsnm{Necula}, \binits{M.}},
\bauthor{\bsnm{Petcu}, \binits{D.}},
\bauthor{\bsnm{Stefan}, \binits{A.}}:
\batitle{Natural language processing in requirements engineering: A systematic mapping study}.
\bjtitle{Electronics}
\bvolume{13}(\bissue{11}),
\bfpage{2055}
(\byear{2024})
\doiurl{10.3390/electronics13112055}
\end{barticle}
\endbibitem

\bibitem[\protect\citeauthoryear{Al-Shalif et~al.}{2024}]{alshalif2024}
\begin{barticle}
\bauthor{\bsnm{Al-Shalif}, \binits{A.}},
\bauthor{\bsnm{Aljarah}, \binits{I.}},
\bauthor{\bsnm{Mirjalili}, \binits{S.}}:
\batitle{A comprehensive review of metaheuristic-based feature selection for text classification}.
\bjtitle{PeerJ Computer Science}
\bvolume{10},
\bfpage{2084}
(\byear{2024})
\doiurl{10.7717/peerj-cs.2084}
\end{barticle}
\endbibitem

\bibitem[\protect\citeauthoryear{Anghelescu et~al.}{2023}]{anghelescu2023}
\begin{barticle}
\bauthor{\bsnm{Anghelescu}, \binits{A.}},
\bauthor{\bsnm{Gheorghe}, \binits{G.}},
\bauthor{\bsnm{Suciu}, \binits{B.A.}},
\bauthor{\bsnm{Suciu}, \binits{G.}}:
\batitle{Chatgpt utility in healthcare education, research, and practice: Systematic review on the promising perspectives and valid concerns}.
\bjtitle{Balneo and PRM Research Journal}
\bvolume{14}(\bissue{4}),
\bfpage{1}--\blpage{9}
(\byear{2023})
\doiurl{10.12680/balneo.2023.614}
\end{barticle}
\endbibitem

\bibitem[\protect\citeauthoryear{Mahuli et~al.}{2023}]{mahuli2023}
\begin{barticle}
\bauthor{\bsnm{Mahuli}, \binits{S.A.}},
\bauthor{\bsnm{Rai}, \binits{A.}},
\bauthor{\bsnm{Mahuli}, \binits{A.V.}},
\bauthor{\bsnm{Kumar}, \binits{A.}}:
\batitle{Application chatgpt in conducting systematic reviews and meta-analyses}.
\bjtitle{British Dental Journal}
\bvolume{235}(\bissue{2}),
\bfpage{90}--\blpage{92}
(\byear{2023})
\doiurl{10.1038/s41415-023-6132-y}
\end{barticle}
\endbibitem

\bibitem[\protect\citeauthoryear{Najafali et~al.}{2023}]{najafali2023}
\begin{barticle}
\bauthor{\bsnm{Najafali}, \binits{D.}},
\bauthor{\bsnm{Camacho}, \binits{J.M.}},
\bauthor{\bsnm{Reiche}, \binits{E.}},
\bauthor{\bsnm{Galbraith}, \binits{L.G.}},
\bauthor{\bsnm{Morrison}, \binits{S.D.}},
\bauthor{\bsnm{Dorafshar}, \binits{A.H.}}:
\batitle{Truth or lies? the pitfalls and limitations of chatgpt in systematic review creation}.
\bjtitle{Aesthetic Surgery Journal}
\bvolume{43}(\bissue{8}),
\bfpage{654}--\blpage{655}
(\byear{2023})
\doiurl{10.1093/asj/sjad093}
\end{barticle}
\endbibitem

\bibitem[\protect\citeauthoryear{Qureshi et~al.}{2023}]{qureshi2023}
\begin{barticle}
\bauthor{\bsnm{Qureshi}, \binits{R.}},
\bauthor{\bsnm{Shaughnessy}, \binits{D.}},
\bauthor{\bsnm{Gill}, \binits{K.A.R.}},
\bauthor{\bsnm{Robinson}, \binits{K.A.}},
\bauthor{\bsnm{Li}, \binits{T.}},
\bauthor{\bsnm{Agai}, \binits{E.}}:
\batitle{Are chatgpt and large language models “the answer” to bringing us closer to systematic review automation?}
\bjtitle{Systematic Reviews}
\bvolume{12}(\bissue{1}),
\bfpage{72}
(\byear{2023})
\doiurl{10.1186/s13643-023-02243-z}
\end{barticle}
\endbibitem

\bibitem[\protect\citeauthoryear{Huotala et~al.}{2024}]{huotala2024}
\begin{bchapter}
\bauthor{\bsnm{Huotala}, \binits{A.}},
\bauthor{\bsnm{Kuutila}, \binits{M.}},
\bauthor{\bsnm{Ralph}, \binits{P.}},
\bauthor{\bsnm{M\"{a}ntyl\"{a}}, \binits{M.}}:
\bctitle{The promise and challenges of using llms to accelerate the screening process of systematic reviews}.
In: \bbtitle{Proceedings of the 28th International Conference on Evaluation and Assessment in Software Engineering}.
\bsertitle{EASE '24},
pp. \bfpage{262}--\blpage{271}.
\bpublisher{Association for Computing Machinery},
\blocation{New York, NY, USA}
(\byear{2024}).
\doiurl{10.1145/3661167.3661172}
\end{bchapter}
\endbibitem

\bibitem[\protect\citeauthoryear{Felizardo et~al.}{2024}]{felizardo2024}
\begin{bchapter}
\bauthor{\bsnm{Felizardo}, \binits{K.R.}},
\bauthor{\bsnm{Lima}, \binits{M.S.}},
\bauthor{\bsnm{Deizepe}, \binits{A.}},
\bauthor{\bsnm{Conte}, \binits{T.U.}},
\bauthor{\bsnm{Steinmacher}, \binits{I.}}:
\bctitle{Chatgpt application in systematic literature reviews in software engineering: an evaluation of its accuracy to support the selection activity}.
In: \bbtitle{Proceedings of the 18th ACM / IEEE International Symposium on Empirical Software Engineering and Measurement (ESEM '24)}
(\byear{2024}).
\doiurl{10.1145/3674805.3686666}
\end{bchapter}
\endbibitem

\bibitem[\protect\citeauthoryear{Dang et~al.}{2022}]{dang2022_ArXiv_Prompts}
\begin{botherref}
\oauthor{\bsnm{Dang}, \binits{H.}},
\oauthor{\bsnm{Mecke}, \binits{L.}},
\oauthor{\bsnm{Lehmann}, \binits{F.}},
\oauthor{\bsnm{Goller}, \binits{S.}},
\oauthor{\bsnm{Buschek}, \binits{D.}}:
How to Prompt? Opportunities and Challenges of Zero- and Few-Shot Learning for Human-AI Interaction in Creative Applications of Generative Models
(2022).
\url{https://arxiv.org/abs/2209.01390}
\end{botherref}
\endbibitem

\bibitem[\protect\citeauthoryear{Sahoo et~al.}{2025}]{sahoo2025_ARXiv_SSPE}
\begin{botherref}
\oauthor{\bsnm{Sahoo}, \binits{P.}},
\oauthor{\bsnm{Singh}, \binits{A.K.}},
\oauthor{\bsnm{Saha}, \binits{S.}},
\oauthor{\bsnm{Jain}, \binits{V.}},
\oauthor{\bsnm{Mondal}, \binits{S.}},
\oauthor{\bsnm{Chadha}, \binits{A.}}:
A Systematic Survey of Prompt Engineering in Large Language Models: Techniques and Applications
(2025).
\url{https://arxiv.org/abs/2402.07927}
\end{botherref}
\endbibitem

\bibitem[\protect\citeauthoryear{Liu et~al.}{2023}]{Liu_2023_PromptingMethods}
\begin{botherref}
\oauthor{\bsnm{Liu}, \binits{P.}},
\oauthor{\bsnm{Yuan}, \binits{W.}},
\oauthor{\bsnm{Fu}, \binits{J.}},
\oauthor{\bsnm{Jiang}, \binits{Z.}},
\oauthor{\bsnm{Hayashi}, \binits{H.}},
\oauthor{\bsnm{Neubig}, \binits{G.}}:
Pre-train, prompt, and predict: A systematic survey of prompting methods in natural language processing
\textbf{55}(9)
(2023)
\doiurl{10.1145/3560815}
\end{botherref}
\endbibitem

\bibitem[\protect\citeauthoryear{Marvin et~al.}{2024}]{Marvin_2023_Prompt}
\begin{bchapter}
\bauthor{\bsnm{Marvin}, \binits{G.}},
\bauthor{\bsnm{Hellen}, \binits{N.}},
\bauthor{\bsnm{Jjingo}, \binits{D.}},
\bauthor{\bsnm{Nakatumba-Nabende}, \binits{J.}}:
\bctitle{Prompt engineering in large language models}.
In: \beditor{\bsnm{Jacob}, \binits{I.J.}},
\beditor{\bsnm{Piramuthu}, \binits{S.}},
\beditor{\bsnm{Falkowski-Gilski}, \binits{P.}} (eds.)
\bbtitle{Data Intelligence and Cognitive Informatics},
pp. \bfpage{387}--\blpage{402}.
\bpublisher{Springer},
\blocation{Singapore}
(\byear{2024}).
\doiurl{10.1007/978-981-99-7962-2_30}
\end{bchapter}
\endbibitem

\bibitem[\protect\citeauthoryear{Xu et~al.}{2025}]{xu2025_expertprompting}
\begin{botherref}
\oauthor{\bsnm{Xu}, \binits{B.}},
\oauthor{\bsnm{Yang}, \binits{A.}},
\oauthor{\bsnm{Lin}, \binits{J.}},
\oauthor{\bsnm{Wang}, \binits{Q.}},
\oauthor{\bsnm{Zhou}, \binits{C.}},
\oauthor{\bsnm{Zhang}, \binits{Y.}},
\oauthor{\bsnm{Mao}, \binits{Z.}}:
ExpertPrompting: Instructing Large Language Models to be Distinguished Experts
(2025).
\url{https://arxiv.org/abs/2305.14688}
\end{botherref}
\endbibitem

\bibitem[\protect\citeauthoryear{Chen et~al.}{2023}]{Chen_2023_Unleashing_prompt}
\begin{barticle}
\bauthor{\bsnm{Chen}, \binits{B.}},
\bauthor{\bsnm{Zhang}, \binits{Z.}},
\bauthor{\bsnm{Langren{\'e}}, \binits{N.}},
\bauthor{\bsnm{Zhu}, \binits{S.}}:
\batitle{Unleashing the potential of prompt engineering in large language models: a comprehensive review}.
\bjtitle{arXiv preprint arXiv:2310.14735}
(\byear{2023})
\doiurl{10.48550/arXiv.2310.14735}
\end{barticle}
\endbibitem

\bibitem[\protect\citeauthoryear{Son et~al.}{2025}]{Minjun_2025_prompt}
\begin{botherref}
\oauthor{\bsnm{Son}, \binits{M.}},
\oauthor{\bsnm{Won}, \binits{Y.-J.}},
\oauthor{\bsnm{Lee}, \binits{S.}}:
Optimizing large language models: A deep dive into effective prompt engineering techniques.
Applied Sciences
\textbf{15}(3)
(2025)
\doiurl{10.3390/app15031430}
\end{botherref}
\endbibitem

\bibitem[\protect\citeauthoryear{Ekin}{2023}]{Sabit_2023_PE4ChatGpt}
\begin{botherref}
\oauthor{\bsnm{Ekin}, \binits{S.}}:
Prompt Engineering For ChatGPT: A Quick Guide To Techniques, Tips, And Best Practices.
Institute of Electrical and Electronics Engineers (IEEE)
(2023).
\doiurl{10.36227/techrxiv.22683919.v1} .
\url{http://dx.doi.org/10.36227/techrxiv.22683919.v1}
\end{botherref}
\endbibitem

\bibitem[\protect\citeauthoryear{Clarisó and Cabot}{2023}]{Clariso_Cabot_2023}
\begin{bchapter}
\bauthor{\bsnm{Clarisó}, \binits{R.}},
\bauthor{\bsnm{Cabot}, \binits{J.}}:
\bctitle{Model-driven prompt engineering}.
In: \bbtitle{2023 ACM/IEEE 26th International Conference on Model Driven Engineering Languages and Systems (MODELS)},
pp. \bfpage{47}--\blpage{54}
(\byear{2023}).
\doiurl{10.1109/MODELS58315.2023.00020}
\end{bchapter}
\endbibitem

\bibitem[\protect\citeauthoryear{Tonmoy et~al.}{2024}]{Tonmoy_2024comprehensive_allucinations}
\begin{botherref}
\oauthor{\bsnm{Tonmoy}, \binits{S.}},
\oauthor{\bsnm{Zaman}, \binits{S.}},
\oauthor{\bsnm{Jain}, \binits{V.}},
\oauthor{\bsnm{Rani}, \binits{A.}},
\oauthor{\bsnm{Rawte}, \binits{V.}},
\oauthor{\bsnm{Chadha}, \binits{A.}},
\oauthor{\bsnm{Das}, \binits{A.}}:
A comprehensive survey of hallucination mitigation techniques in large language models.
arXiv preprint arXiv:2401.01313
\textbf{6}
(2024)
\end{botherref}
\endbibitem

\bibitem[\protect\citeauthoryear{Lewis et~al.}{2020}]{Lewis_2020_RAG}
\begin{bchapter}
\bauthor{\bsnm{Lewis}, \binits{P.}},
\bauthor{\bsnm{Perez}, \binits{E.}},
\bauthor{\bsnm{Piktus}, \binits{A.}},
\bauthor{\bsnm{Petroni}, \binits{F.}},
\bauthor{\bsnm{Karpukhin}, \binits{V.}},
\bauthor{\bsnm{Goyal}, \binits{N.}},
\bauthor{\bsnm{K\"{u}ttler}, \binits{H.}},
\bauthor{\bsnm{Lewis}, \binits{M.}},
\bauthor{\bsnm{Yih}, \binits{W.-t.}},
\bauthor{\bsnm{Rockt\"{a}schel}, \binits{T.}},
\bauthor{\bsnm{Riedel}, \binits{S.}},
\bauthor{\bsnm{Kiela}, \binits{D.}}:
\bctitle{Retrieval-augmented generation for knowledge-intensive nlp tasks}.
In: \bbtitle{Proceedings of the 34th International Conference on Neural Information Processing Systems}.
\bsertitle{NIPS '20}.
\bpublisher{Curran Associates Inc.},
\blocation{Red Hook, NY, USA}
(\byear{2020})
\end{bchapter}
\endbibitem

\bibitem[\protect\citeauthoryear{Cohen}{1960}]{Cohen_Kappa}
\begin{barticle}
\bauthor{\bsnm{Cohen}, \binits{J.}}:
\batitle{A coefficient of agreement for nominal scales}.
\bjtitle{Educational and Psychological Measurement}
\bvolume{20}(\bissue{1}),
\bfpage{37}--\blpage{46}
(\byear{1960})
\doiurl{10.1177/001316446002000104}
\end{barticle}
\endbibitem

\bibitem[\protect\citeauthoryear{Fleiss}{1971}]{fleiss_Kappa}
\begin{barticle}
\bauthor{\bsnm{Fleiss}, \binits{J.L.}}:
\batitle{Measuring nominal scale agreement among many raters}.
\bjtitle{Psychological Bulletin}
\bvolume{76}(\bissue{5}),
\bfpage{378}--\blpage{382}
(\byear{1971})
\doiurl{10.1037/h0031619}
\end{barticle}
\endbibitem

\bibitem[\protect\citeauthoryear{Fletcher and Fletcher}{2005}]{Fletcher2005_PPA}
\begin{bbook}
\bauthor{\bsnm{Fletcher}, \binits{R.H.}},
\bauthor{\bsnm{Fletcher}, \binits{S.W.}}:
\bbtitle{Clinical Epidemiology},
\bedition{4}th edn.
\bpublisher{Lippincott Williams and Wilkins},
\blocation{ISBN 0-7817-5215-9. Philadelphia, PA}
(\byear{2005})
\end{bbook}
\endbibitem

\bibitem[\protect\citeauthoryear{Dubinin et~al.}{2021}]{CaptchaPaper_Dubinin_2021}
\begin{barticle}
\bauthor{\bsnm{Dubinin}, \binits{D.V.}},
\bauthor{\bsnm{Kochegurov}, \binits{A.I.}},
\bauthor{\bsnm{Geringer}, \binits{V.E.}}:
\batitle{Improving the criteria for quality assessment of image processing algorithms}.
\bjtitle{Journal of Physics: Conference Series}
\bvolume{1862}(\bissue{1}),
\bfpage{012011}
(\byear{2021})
\doiurl{10.1088/1742-6596/1862/1/012011}
\end{barticle}
\endbibitem

\end{thebibliography}

\end{document}